%% file: ms.tex
\documentclass[twocolumn]{aastex63}

\newcommand{\teff}{T_{\rm eff}} 
\newcommand{\gr}{g\, -\, r} 
\newcommand{\gi}{g\, -\, i} 
\newcommand{\gz}{g\, -\, z} 
\newcommand{\ug}{u\, -\, g}

\accepted{for publication in the Astrophysical Journal}

\shorttitle{A Blueprint for the Milky Way's Stellar Populations.\ II.} 
\shortauthors{An \& Beers}

\begin{document}

\title{A Blueprint for the Milky Way's Stellar Populations.\ II. Improved Isochrone Calibration in the SDSS and Pan-STARRS Photometric Systems}

\correspondingauthor{Deokkeun An} 
\email{deokkeun@ewha.ac.kr}

\author{Deokkeun An} 
\affiliation{Department of Science Education, Ewha Womans University, Seoul 03760, Republic of Korea}

\author{Timothy C.\ Beers} 
\affiliation{Department of Physics and JINA Center for the Evolution of the Elements, University of Notre Dame, Notre Dame, IN 46556, USA}

\begin{abstract}

We improve the identification and isolation of individual stellar populations in the Galactic halo based on an updated set of empirically calibrated stellar isochrones in the Sloan Digital Sky Survey (SDSS) and Pan-STARRS~1 (PS1) photometric systems. Along the Galactic prime meridian ($l=0^{\circ}$ and $180^{\circ}$), where proper motions and parallaxes from {\it Gaia} DR2 can be used to compute rotational velocities of stars in the rest frame of the Milky Way, we use the observed double color-magnitude sequences of stars having large transverse motions, which are attributed to groups of stars in the metal-poor halo and the thick disk with halo-like kinematics, respectively. The {\it Gaia} sequences directly constrain color-magnitude relations of model colors, and help to improve our previous calibration using Galactic star clusters. Based on these updated sets of stellar isochrones, we confirm earlier results on the presence of distinct groups of stars in the metallicity versus rotational-velocity plane, and find that the distribution of the most metal-poor ([Fe/H] $<-2$) stars in our sample can be modeled using two separate groups on prograde and retrograde orbits, respectively. At $4$--$6$~kpc from the Galactic plane, we find approximately equal proportions of the Splashed Disk, and the metal-rich ($\langle {\rm [Fe/H]} \rangle\sim-1.6$) and metal-poor ($\langle {\rm [Fe/H]} \rangle\sim-2.5$) halos on prograde orbits. The {\it Gaia}-Sausage-Enceladus, the metal-weak thick disk, and the retrograde halo structure(s) ($\langle {\rm [Fe/H]} \rangle\sim-2.2$) constitute approximately $10\%$ of the rest of the stellar populations at these distances.

\end{abstract}


\section{Introduction}

Color-magnitude relations of stars are basic tools for studying the fundamental properties of stellar systems. Nevertheless, theoretical approaches exhibit limitations in matching model predictions to observed sequences of well-studied star clusters. This mismatch largely arises from an inaccurate conversion from theoretically predicted effective temperatures ($\teff$) to observable colors, which reflects the complex nature and interplay of stellar interior and atmosphere models \citep[e.g., ][]{pinsonneault:03}. To overcome this situation, semi-empirical modeling has been suggested as a practical means for correcting theoretically predicted quantities \citep[e.g., ][]{lejeune:97,lejeune:98,westera:02, vandenberg:03,pinsonneault:04, an:09b,an:13,an:15b}. Stellar isochrones calibrated in his way have been successfully used for deriving accurate distances and metallicities of individual stars and star clusters \citep[e.g.,][]{an:07,an:09a,an:13, an:15a,an:19,an:20}.

Ideally, samples for the calibration of stellar models should comprise a single stellar population with well-defined physical parameters such as distance and chemical compositions. For metal-poor stars, the traditional approach is to employ subdwarf stars in the Solar Neighborhood with accurate parallaxes \citep[e.g., {\it Hipparcos}; ][]{esa:97}, but this approach is limited by the small size of the calibration sample. Calibration based on star clusters is also useful, because they provide well-defined sequences for a wide range of stellar masses and luminosity. Some of them also place useful constraints on the physical parameters, although such clusters with good distance measurements are limited to relatively few, nearby systems.

\citet{an:09b} provided an empirically calibrated set of isochrones in the $ugriz$ photometric system \citep{fukugita:96} of the Sloan Digital Sky Survey \citep[SDSS;][]{abolfathi:18}, based on observations of Galactic star clusters. However, there remain some limitations in this study. First, the original cluster photometry in \citet{an:08}, which is employed in the above study, is limited to $\sim2\%$--$3\%$ errors, because putting the crowded-field photometry onto the global scale of the SDSS system can inevitably introduce systematic zero-point errors. In an effort to reduce the size of these errors, \citet{an:13} updated models by matching the \citet{an:08} photometry with the `uber'-calibrated SDSS photometry \citep{padmanabhan:08} in the cluster's (low-density) flanking fields.  Nevertheless, the exact size of the systematic error in the cluster photometry is still poorly constrained.

Secondly, the SDSS imaging data do not cover faint main-sequence (MS) stars in the bright cluster samples, due to the relatively bright completeness limit in the survey -- the $95\%$ completeness limits for point sources are $u=22.0$, $g=22.2$, $r=22.2$, $i=21.3$, and $z=20.5$.  Useful photometry is limited to those of brighter stars, resulting in an effective absolute magnitude limit of $M_r \sim 7$ for the calibration in $\gr$ and $\gi$. For metal-poor halo stars, this limiting magnitude corresponds to masses of $\sim0.6\ M_\odot$. When the color indices include the $u$ or $z$ passbands, the magnitude limit for the model calibration becomes brighter ($M_r \sim 6$ or $0.7\ M_\odot$). This inhomogeneous coverage in stellar mass inevitably limits the application of the isochrones, which in many cases requires multi-band photometry to constrain stellar parameters.

Thirdly, the cluster-based calibration relies on subdwarf-fitting distances to globular clusters based on {\it Hipparcos} parallaxes of nearby subdwarfs \citep[e.g.,][]{reid:97,carretta:00,kraft:03}. Such distance determinations typically have $\sim5\%$--$7\%$ errors. If the luminosity of a star is taken as an independent variable to compute color differences between observation and the model \citep{pinsonneault:04,an:09b}, the current uncertainty in the cluster distances translates into a moderate systematic offset in colors, on the order of $0.02$--$0.04$~mag.

Lastly, the globular cluster samples with both good cluster parameters (distance, reddening, and metallicity [Fe/H]\footnote{Throughout this paper, we use [Fe/H] to indicate a logarithmic iron abundance with respect to the solar value. Since we employ models with a specific relation between [Fe/H] and $\alpha$-elemental abundance ([$\alpha$/Fe]) of stars (see \S~\ref{sec:model}), an overall metallicity ([M/H]) is higher than [Fe/H] for metal-poor stars with elevated [$\alpha$/Fe].}) and accurate crowded-field photometry are limited to those with [Fe/H] $\la-1.2$, while the open cluster samples have [Fe/H] $\geq0$. There are metal-rich globular clusters that might be considered, such as 47~Tuc or M71 ([Fe/H] $\sim-0.7$), but they are either too distant or attenuated by a large amount of foreground dust. None of the other clusters within the metallicity gap have well-constrained parameters and/or accurate $ugriz$ photometry.

To overcome the above limitations, we turn our attention to a dynamical family of stars in the Milky Way as a calibration sample. In particular, we focus on the double sequences in a color-magnitude diagram (CMD) of stars with large transverse motions, discovered from the analysis of {\it Gaia} DR2 \citep{gaiahrd,gaiadr2}. These sequences are ascribed to metal-poor halo stars and thick-disk (TD) stars with halo-like kinematics in the Galaxy, respectively. More clear identification of stars belonging to each sequence can be seen in \citet[][hereafter Paper~I]{an:20}. According to the chemo-dynamical mapping in Paper~I, the blue sequence is composed of stars with $\langle {\rm [Fe/H]} \rangle \sim -1.3$, and is likely a collection of stars from the so-called inner halo \citep[IH;][]{carollo:07,carollo:10,beers:12} and the debris of {\it Gaia}-Sausage-Enceladus \citep[GSE;][]{belokurov:18, helmi:18}. On the other hand, the red sequence has $\langle {\rm [Fe/H]} \rangle \sim -0.4$, and belongs to the Splashed Disk \citep[SD; ][]{bonaca:17,belokurov:20}.

The use of {\it Gaia}'s double sequences can help avoid problems of imprecise distance estimates for the Galactic clusters and the global zero-point variations in the survey data. Stars that constitute {\it Gaia}'s sequences are relatively nearby, and therefore have good proper-motion and parallax measurements, in addition to low foreground reddening. Because of their proximity to the Sun, the observed MS extends far below the faint limit set by the former calibrating samples of Galactic globular clusters, and therefore cover a wider range of stellar mass. Since these stars are spread over the celestial sphere, the observed sequences represent the average color-magnitude relation of each dynamical family, and are less affected by local photometric zero-point errors. In addition, the red sequence has an intermediate metallicity between the globular and open clusters, so it can fill in the missing metallicity baseline in the model calibration.

In this paper, we employ {\it Gaia}'s double sequences to improve our previous calibration of isochrone colors. With the same methodology, we also construct an empirically calibrated set of models in the Pan-STARRS~1 (PS1) photometric system \citep{tonry:12,chambers:16}. This paper is organized as follows. In \S~\ref{sec:sample}, we describe how we extract {\it Gaia}'s double sequences from stars with good astrometric quantities in {\it Gaia} DR2. A summary on the model construction is provided in \S~\ref{sec:model}, followed by a description of the procedures used for the revised empirical corrections in \S~\ref{sec:empcorr}. Updated ``blueprint'' maps showing the metallicity versus rotational-velocity distribution of stars based on these new models and PS1 photometry are presented in \S~\ref{sec:blueprint}.

\begin{figure*}[hbt!] \centering \includegraphics[scale=0.46]{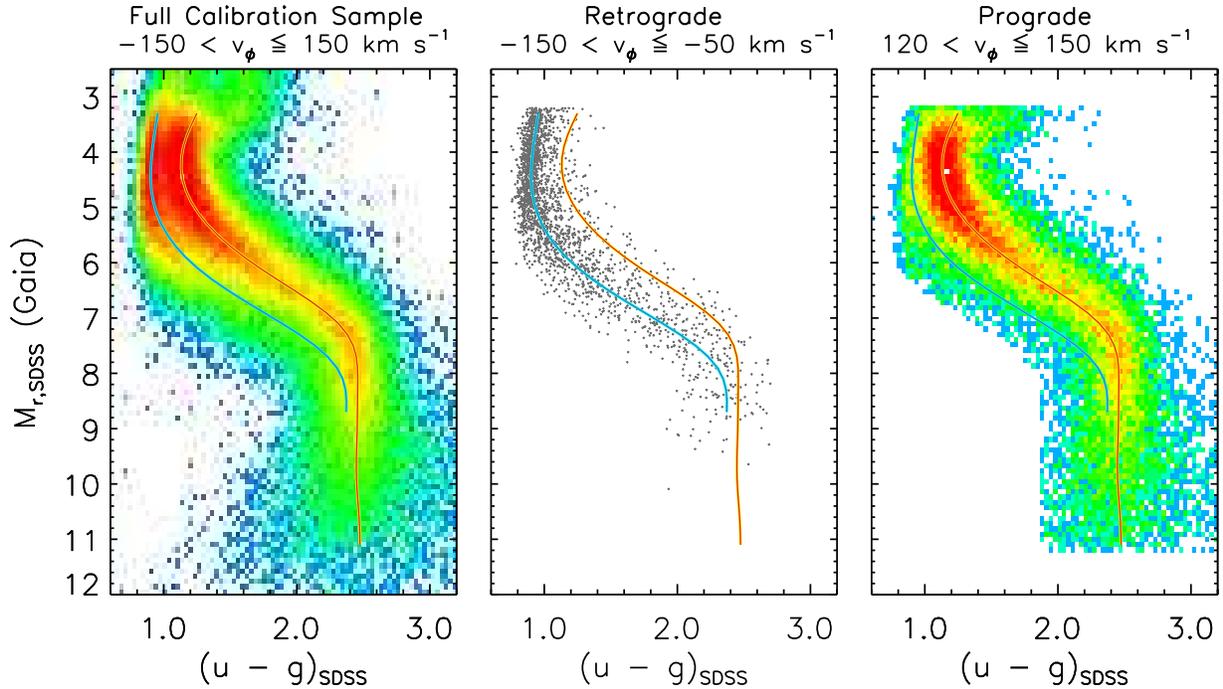} \caption{The $\ug$ color-magnitude diagrams of stars with precise {\it Gaia} parallaxes ($\sigma_\pi/\pi < 0.1$) along the Galactic prime meridian (within $\pm 30\arcdeg$ from $l=0\arcdeg$ and $180\arcdeg$). Rotational velocities in the Galactocentric cylindrical coordinate system are computed using proper-motion and parallax measurements from {\it Gaia} DR2. Left: The number density of all stars with $-150 < v_\phi \leq 150$~km\ s$^{-1}$. Middle: Only stars with large retrograde motions ($-150 < v_\phi \leq -50$~km\ s$^{-1}$). Right: Stars with moderate prograde rotation ($120 < v_\phi \leq 150$~km\ s$^{-1}$). In all panels, the blue and red solid lines indicate fiducial sequences for the retrograde (middle panel) and the prograde (right panel) populations, respectively.} \label{fig:gd} \end{figure*}

\section{Gaia's Double Sequences}\label{sec:sample}

{\it Gaia}'s double sequences are two distinct fiducial lines observed from stars with high transverse motions. In the original work by the {\it Gaia} team, the double sequences are presented in its native photometric filter system, $B_G$ and $R_G$ \citep{gaiahrd}. To utilize these in the isochrone calibration, we reproduce the double sequences in the SDSS \citep{abolfathi:18} and PS1 \citep{chambers:16} photometric systems by cross-matching photometric catalogs with {\it Gaia} DR2.

To refine the sample of stars with high transverse motions, a rotational velocity ($v_\phi$) component in the Galactocentric cylindrical coordinate system is employed, instead of a transverse velocity on the celestial sphere, as used by the {\it Gaia} team. This is physically motivated by our previous work in Paper~I, in which individual stellar components are clearly distinguished from each other in the [Fe/H] versus $v_\phi$ plane. This leads to a more precise isolation of stars that are associated with each of {\it Gaia}'s double sequences. In the absence of radial-velocity measurements for a large number of stars, we compute $v_\phi$ along the Galactic prime meridian ($l=0\arcdeg$ or $180\arcdeg$), where the $v_\phi$ vector essentially depends only on the proper motions and distances (see Paper~I for more information on the $v_\phi$ estimation procedure).

For the calibration sample, only stars with good parallaxes ($\sigma_\pi/\pi < 0.2$) and proper motions [$(\mu_\alpha/\sigma_{\mu_\alpha})^2 + (\mu_\delta/\sigma_{\mu_\delta}) ^2 > (1.0/0.3)^2$] are used. Having good $u$-band photometry ($u < 20.5$) is another important constraint for obtaining accurate fiducial sequences in $\ug$. The foreground extinction estimates in \citet{schlegel:98} are adopted, along with the extinction coefficients of \citet{schlafly:11} in the above filter sets. Since extinction values in \citet{schlegel:98} represent integrated dust absorption along each line of sight, the sample is limited to $|b| > 20\arcdeg$ to minimize the impact of uncertainties in the extinction correction. About $90\%$ of stars in the calibration sample are located above $500$~pc, which is approximately five times the scale height of the Galactic dust layer \citep{drimmel:01,li:18}. A small fraction of stars with $E(B\, -\, V) > 0.1$ are further rejected from the sample. The $E(B\, -\, V)$ distribution of the remaining stars is peaked at $0.02$~mag, with a median value of $0.03$~mag.

\begin{figure*}[hbt!] \centering \includegraphics[scale=0.38]{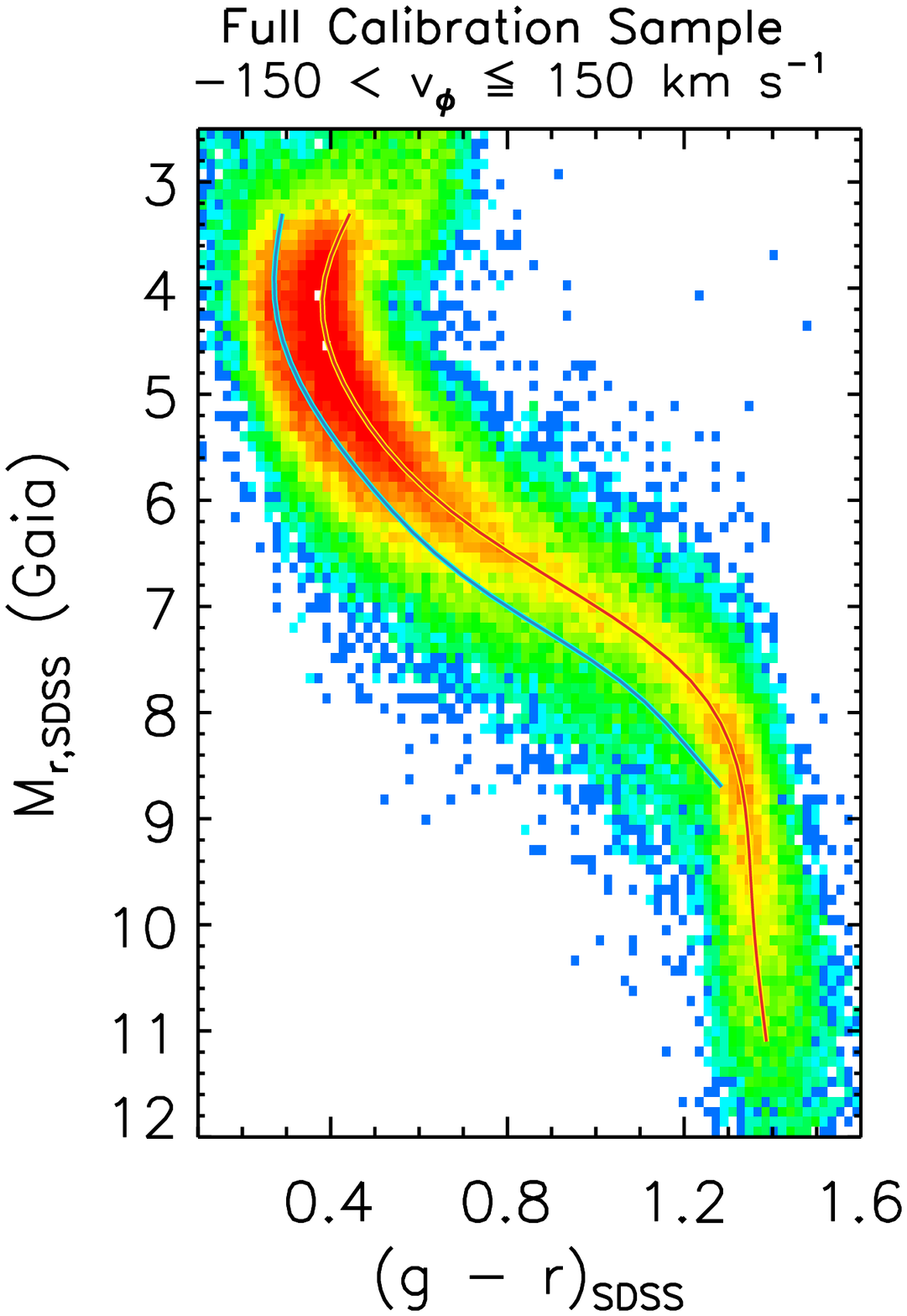} \includegraphics[scale=0.38]{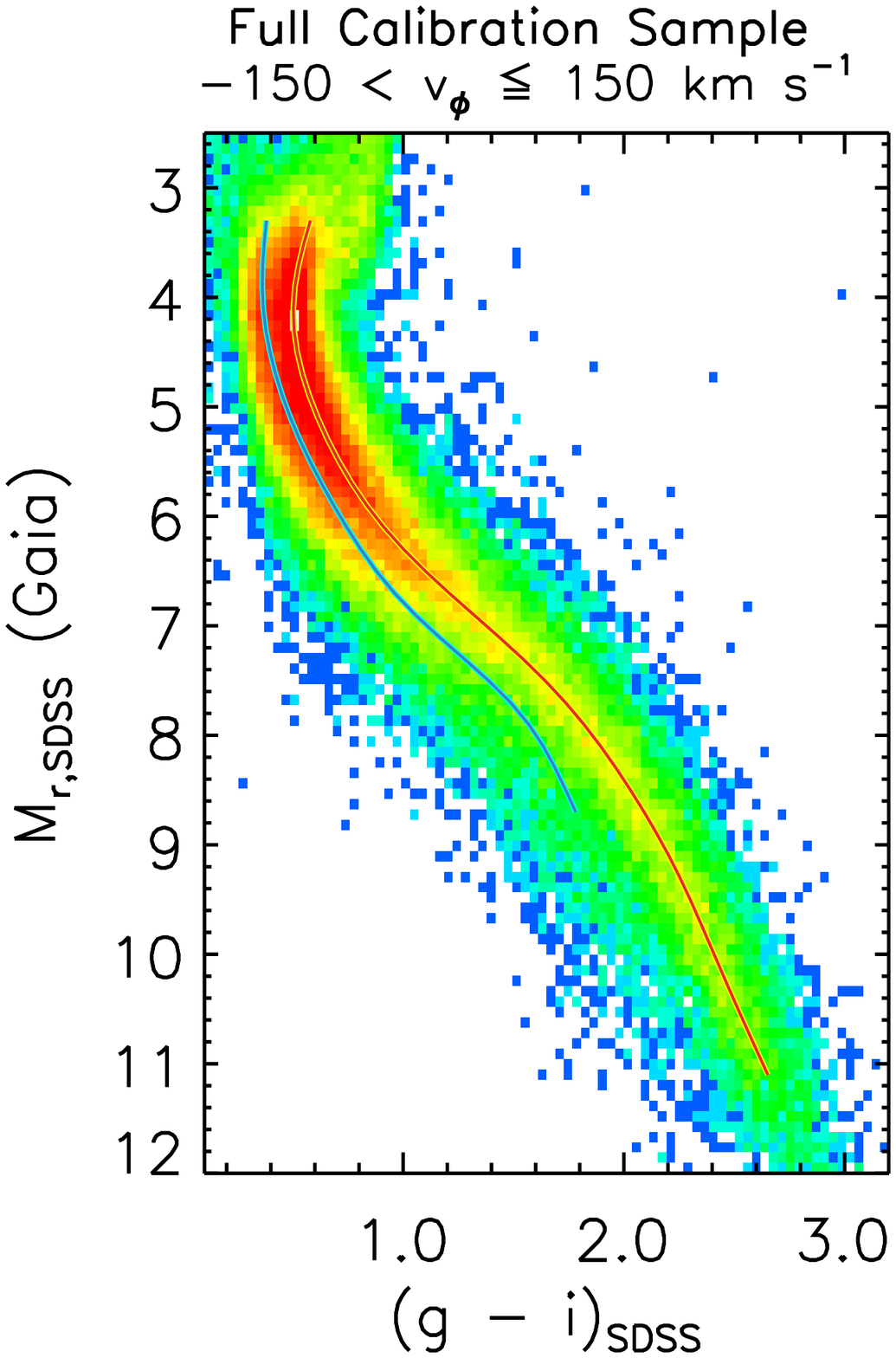} \includegraphics[scale=0.38]{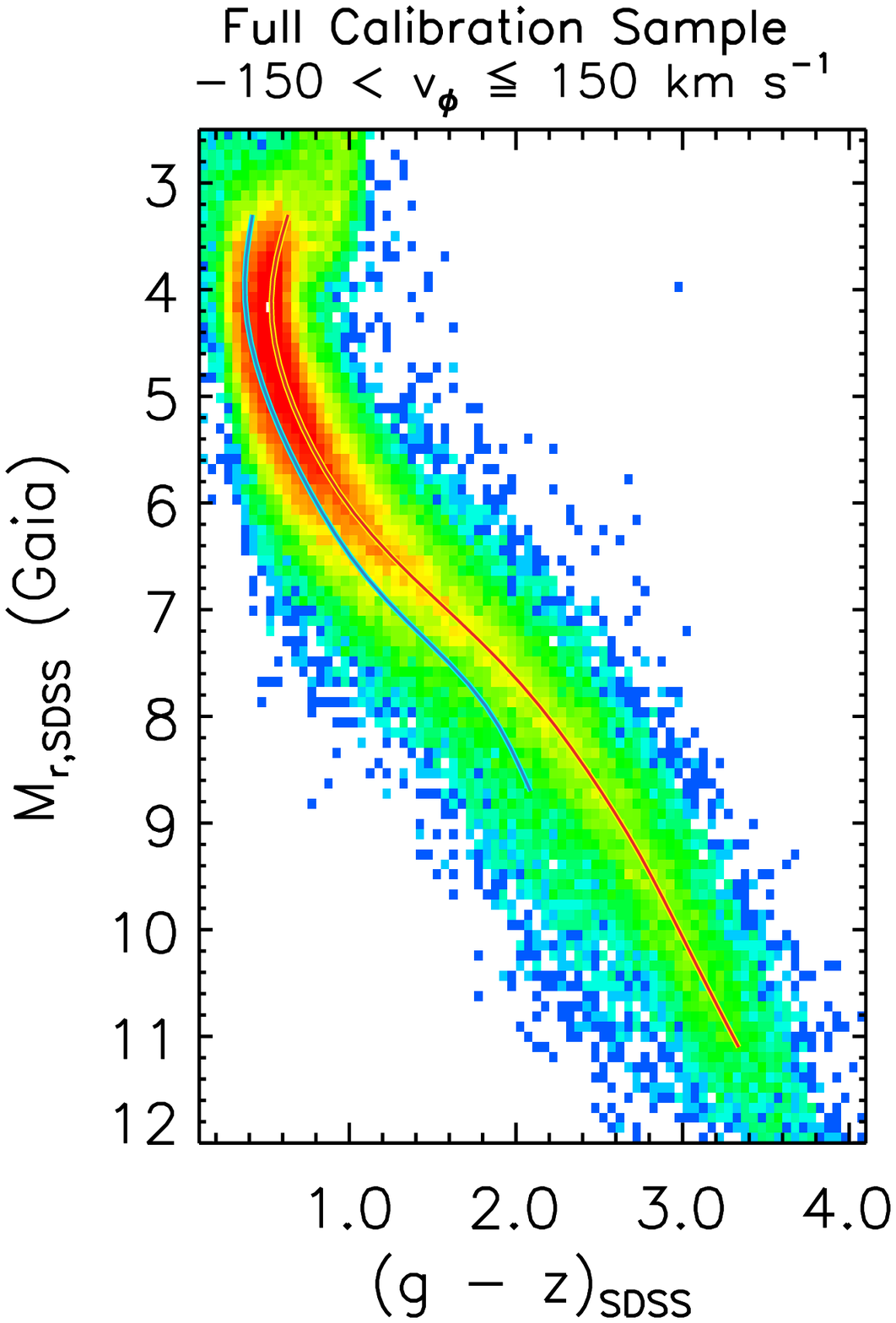} \caption{Same as in the left panel of Figure~\ref{fig:gd}, but displaying CMDs in other color indices.} \label{fig:gd2} \end{figure*}

\begin{figure*}[hbt!] \centering \includegraphics[scale=0.46]{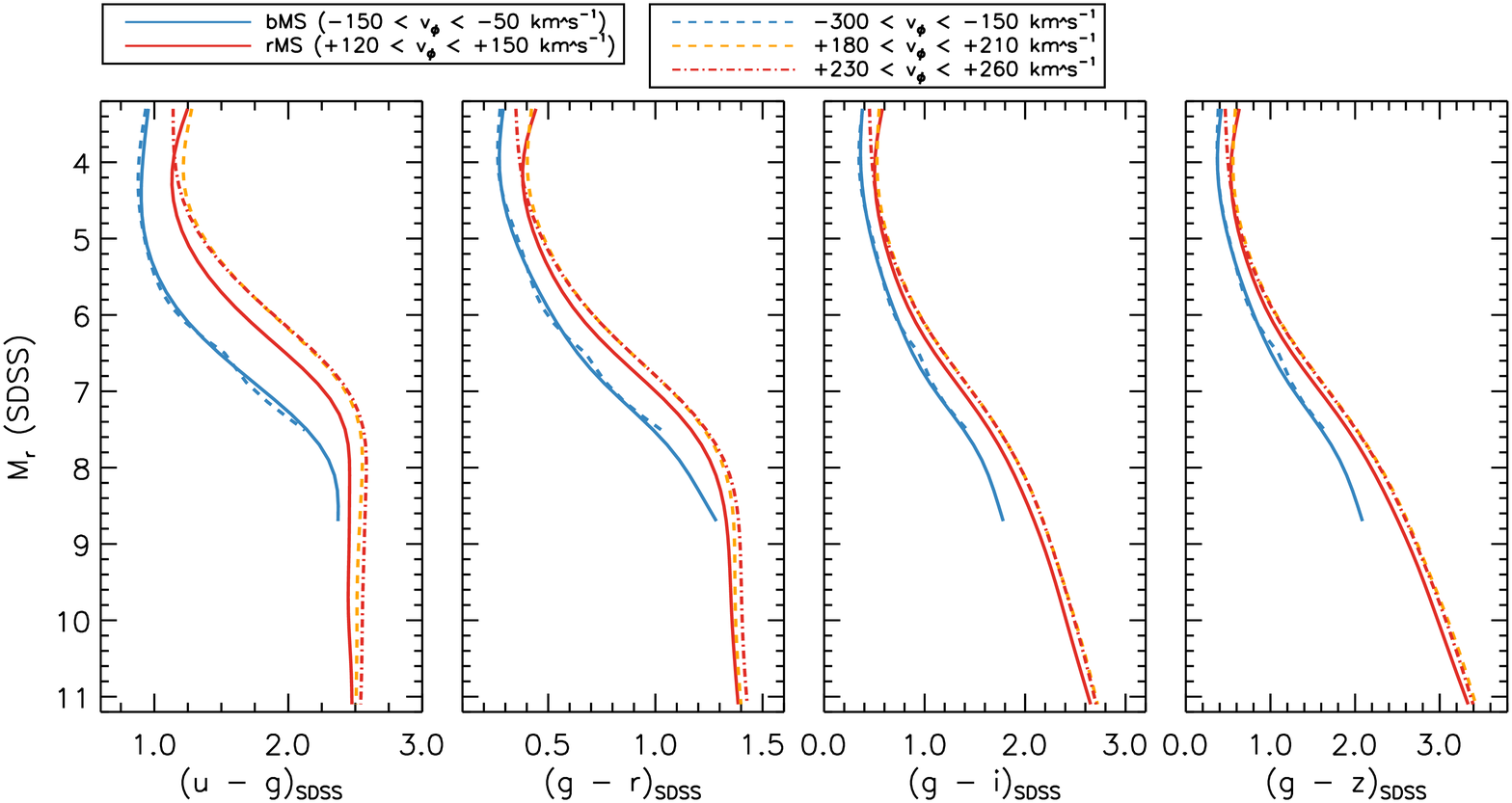} \caption{Fiducial sequences in bins of different rotational velocities.  Each of the lines is obtained based on the same methodology as shown in Figures~\ref{fig:gd}--\ref{fig:gd2}.} \label{fig:gd3} \end{figure*}

The left panel of Figure~\ref{fig:gd} shows the $\ug$ CMD for stars with $-150 < v_\phi < 150$~km~s$^{-1}$, along the Galactic prime meridian within a region of $\pm30\arcdeg$. The computed $v_\phi$ is corrected for a small inclination with respect to the prime meridian (sometimes denoted as the ``projected'' $v_\phi$). The CMDs for the same stars in other color indices ($\gr$, $\gi$, and $\gz$) are displayed in Figure~\ref{fig:gd2}. As shown in Figures~\ref{fig:gd} and \ref{fig:gd2}, the double sequences are most clearly separated in the $\ug$ CMD. The large separation between the blue and the red MS (hereafter bMS and rMS, respectively) is because $\ug$ has the strongest sensitivity to metallicity, and the metallicity difference between the two populations constituting the double sequences is sufficiently large to exhibit a clear division.

\begin{figure*}[hbt!] \centering \includegraphics[scale=0.46]{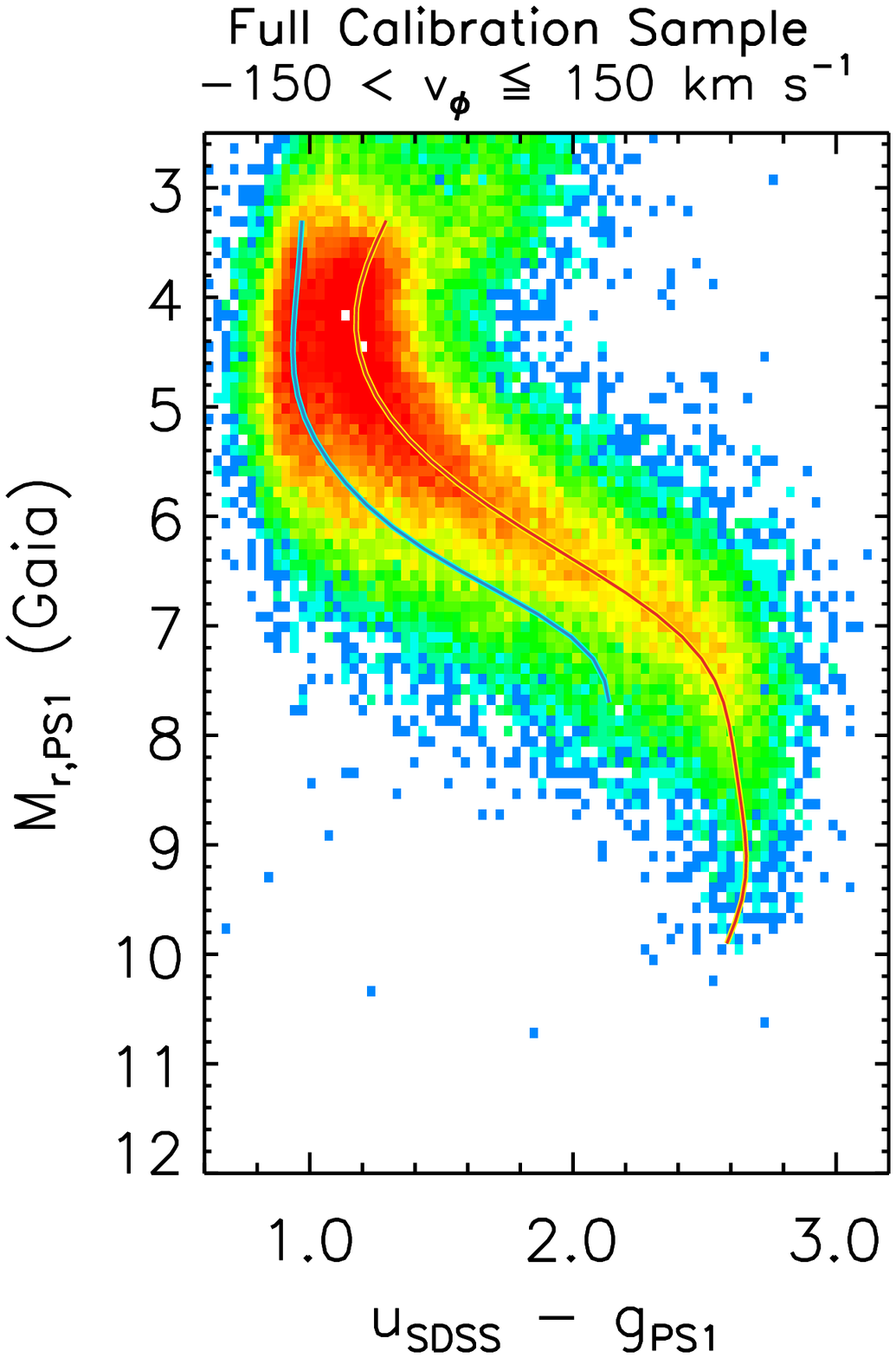} \includegraphics[scale=0.46]{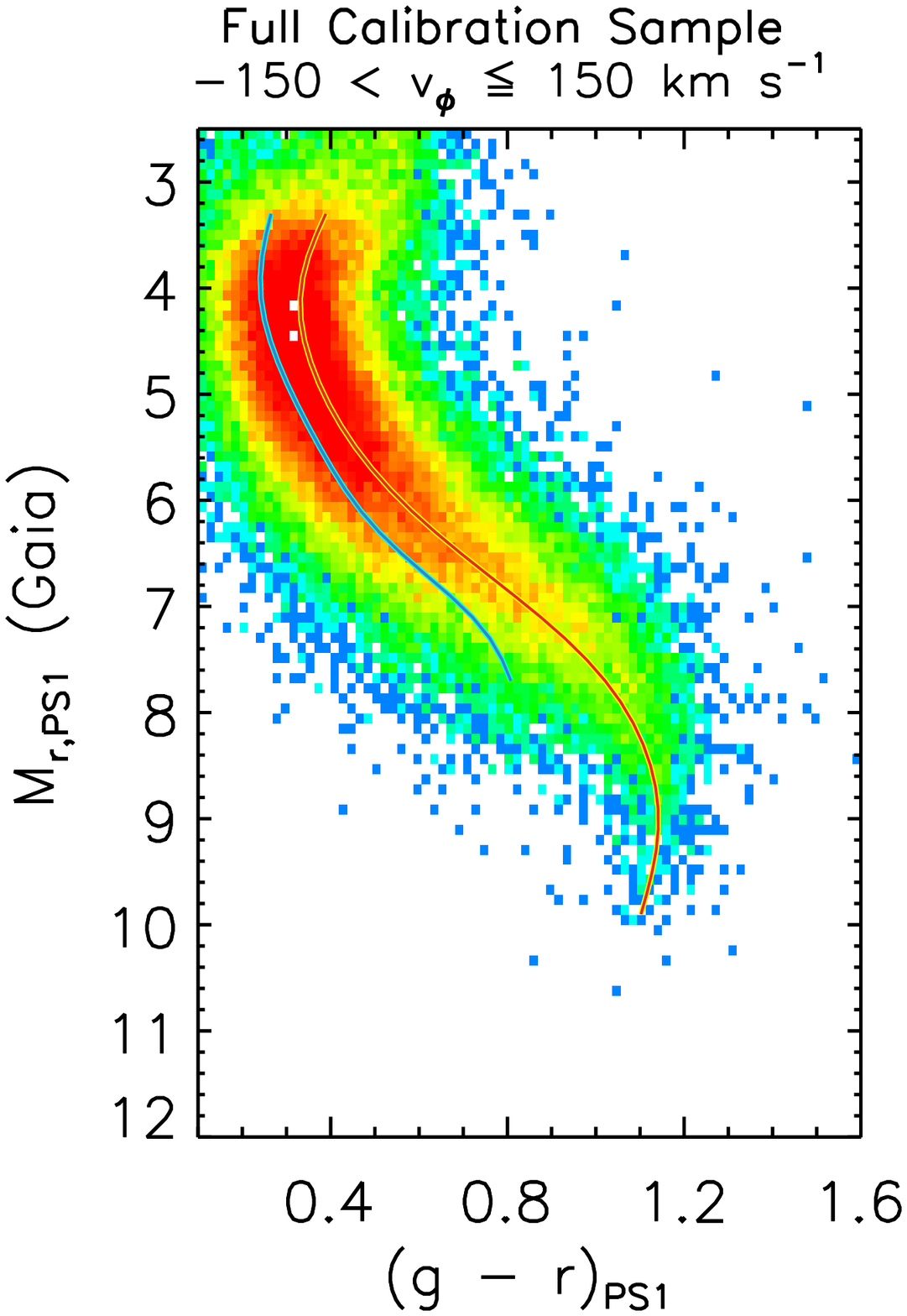} \includegraphics[scale=0.46]{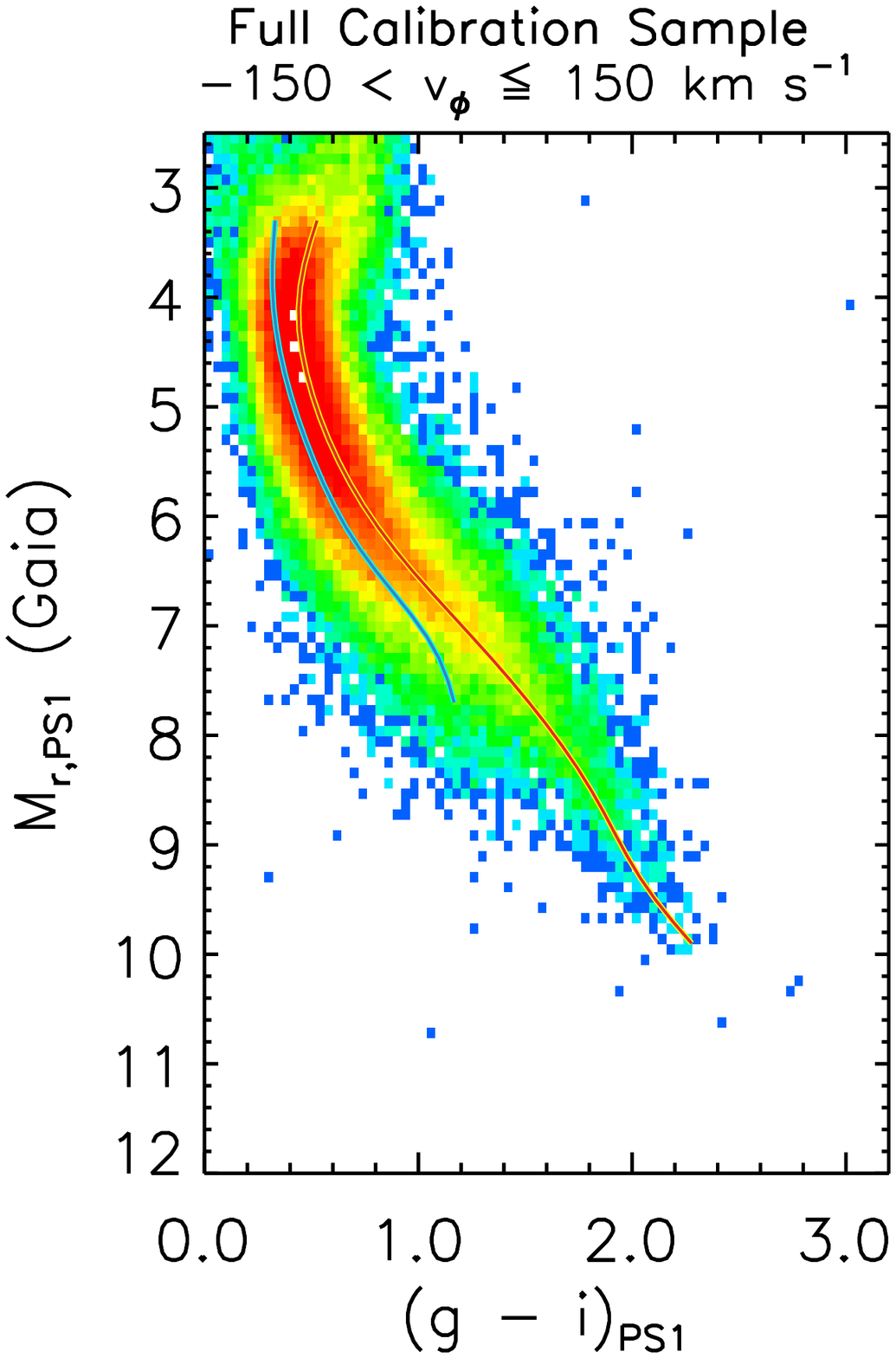} \includegraphics[scale=0.46]{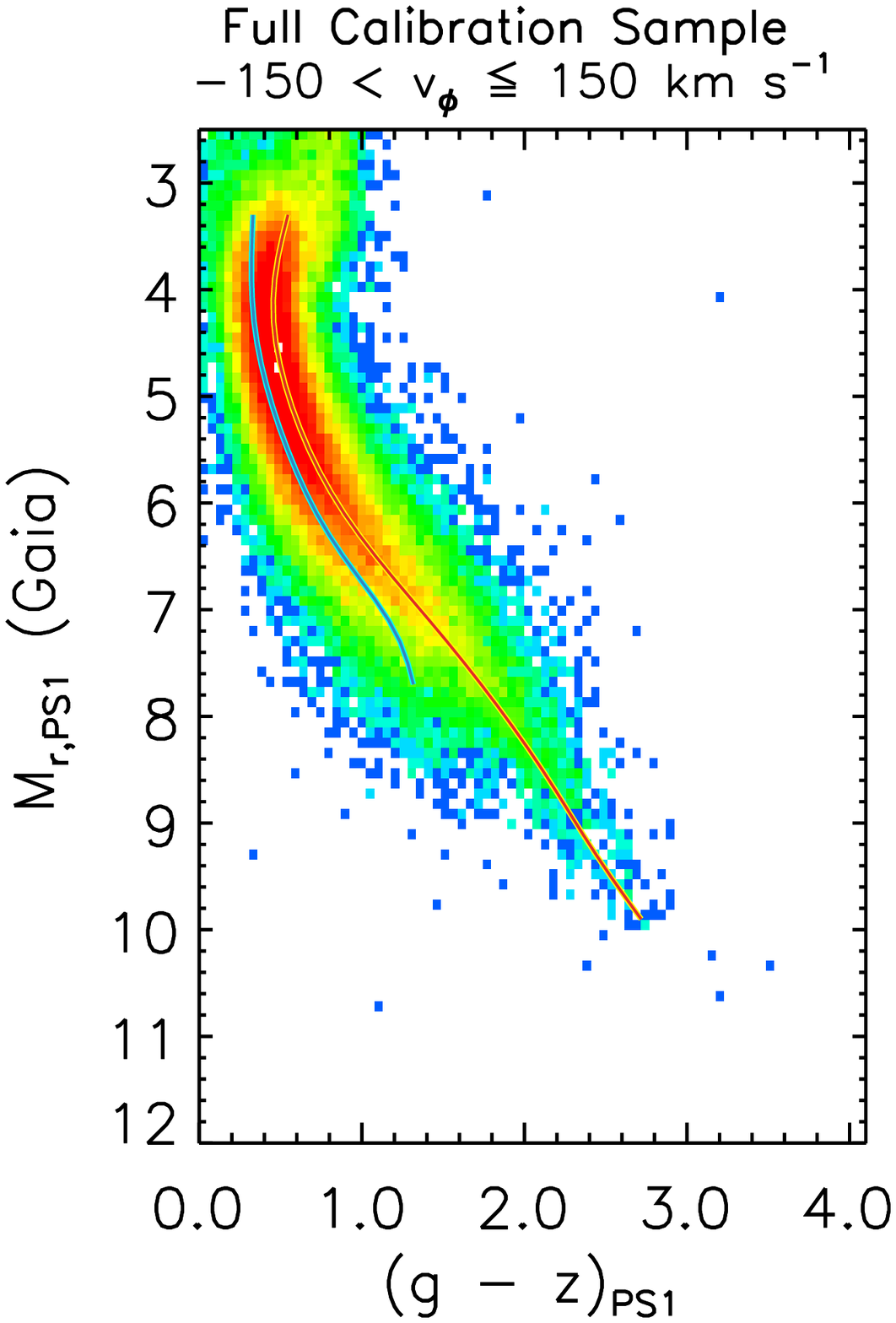} \caption{Same as in the left panel of Figure~\ref{fig:gd}, but displaying CMDs in the PS1 photometric system. The $u$-band data are taken from SDSS.} \label{fig:gdps} \end{figure*}

The blue and red lines in Figures~\ref{fig:gd} and \ref{fig:gd2} indicate fiducial lines for the bMS and rMS, respectively, which trace the median colors of stars in bins of $M_r$. The middle panel in Figure~\ref{fig:gd} includes stars with retrograde motions ($-150 < v_\phi < -50$~km~s$^{-1}$), a subset of the stars shown in the left panel, and the right panel shows stars with prograde motions ($120 < v_\phi < 150$~km~s$^{-1}$). The exact ranges of $v_\phi$ are adjusted until the sequences are most clearly separated from each other, while keeping large numbers of stars in each $v_\phi$ bin. To improve the isolation of stars belonging to each MS, the bMS is constructed after rejecting outliers as those having redder colors than the rMS by $\Delta (\ug) = 0.4$ mag (middle panel). Similarly, the rMS is obtained after rejecting stars bluer than the bMS by $\Delta (\ug) = 0.4$ mag (right panel). This process requires a few iterations; the same groups of stars are used in the derivation of fiducial lines in the other color indices.

\input{tab1.tex} \input{tab2.tex}

The rMS covers a wide range of $M_r$, from the MS turn-off to $M_r\sim11$, which is significantly deeper than the limit set in the best available case of M67. The bMS is shorter than the rMS, due to the smaller number of metal-poor stars in the survey volume. Nevertheless, it is still $\sim3$~mag deeper in $M_r$ than for the bright globular cluster samples in the SDSS photometric catalog \citep{an:08}.

Figure~\ref{fig:gd3} shows a comparison between the bMS and the rMS with fiducial lines obtained using other $v_\phi$ cuts. The fiducial line for stars with $-300 < v_\phi < -150$~km~s$^{-1}$ is similar to the bMS, indicating a similar mean metallicity of these groups of stars. Our previous result in Paper~I indicates that the bMS is mainly composed of stars in the IH and GSE. The contribution of stars from the outer halo \citep[OH;][]{carollo:07,carollo:10} is negligible in local volumes, and therefore the inclusion of stars with higher retrograde velocities makes little change in the derived fiducial line. Regarding metal-rich populations, the rMS includes mostly stars belonging to the SD. For stars with larger $v_\phi$ than $150$~km~s$^{-1}$, their fiducial lines become slightly redder, and the MSTO becomes bluer (younger) than the rMS. The metallicity difference between the two populations is approximately $1$~dex (see below), which helps to bridge the gap between the metal-rich and the metal-poor systems in the model calibration. 

The methodology described above is also applied to the PS1 system (except in the $y$-band). Only the primary detections in the stacked imaging catalog are taken, with a sample cut on $i$-band photometry from a point spread function being less than the $i$-band Kron magnitude plus $0.05$~mag to select point-like sources. Because PS1 does not include $u$-band observations, its $griz$ measurements are cross-matched with the SDSS $u$-band data using a $1\arcsec$ search radius and the same magnitude cut in the $u$-band as above. Figure~\ref{fig:gdps} shows the bMS and the rMS overlaid on CMDs from PS1 (along with the SDSS $u$-band) constructed in this way.

The bMS and rMS in both systems look alike, because of their similar filter-response functions. Nonetheless, they are not identical, and therefore can serve as accurate ``fiducial'' sequences in the native filter systems. The fiducial lines are tabulated in Tables~\ref{tab:tab1} and \ref{tab:tab2} for the SDSS and PS1 systems, respectively.

\section{Theoretical Stellar Isochrones}\label{sec:model}

The same suite of underlying stellar isochrones as in our previous work \citep{an:09b,an:13} is adopted, which is based on the YREC evolutionary models \citep{sills:00} and MARCS atmosphere models \citep{gustafsson:08}. YREC models are generated over a wide range of stellar ages, but the base set consists of $13$~Gyr old models below [Fe/H] $=-1.2$, $4$~Gyr old models above [Fe/H] $=-0.3$, and models at linearly interpolated ages between the two metallicity ranges. The $\teff$ and luminosity predicted by the YREC models are converted into synthetic colors and magnitudes using MARCS atmospheric models. We assume a one-to-one relation between [Fe/H] and $\alpha$-element abundance ([$\alpha$/Fe]) in these models, motivated by spectroscopic observations of stars in the Milky Way \citep[e.g.,][]{venn:04, hayden:15}: [$\alpha$/Fe]$=+0.4$ at [Fe/H] $=-3.0$, [$\alpha$/Fe]$=+0.3$ at $-2.0 \leq {\rm [Fe/H]} \leq -1.0$, [$\alpha$/Fe]$=+0.25$ at [Fe/H] $=-0.75$, [$\alpha$/Fe]$=+0.2$ at [Fe/H] $=-0.5$, and the solar abundance ratio at [Fe/H] $\geq-0.3$. In this study, a new set of synthetic magnitudes in the PS1 photometric bands are computed using filter transmission curves in \citet{tonry:12}. The AB corrections in \citet{abolfathi:18} and \citet{scolnic:15} are adopted in each of the native filter systems.

\section{Derivation of Empirical Color Corrections}\label{sec:empcorr}

In \citet{an:13}, fiducial sequences of M67 are used on the metal-rich side, and those of M92 are taken on the metal-poor side, as a baseline of the calibration. They are partly replaced and reinforced by {\it Gaia}'s double sequences in this revised work. The models are adjusted to match the bMS, along with other globular cluster sequences (M3, M5, M13, M15, and M92), from which color corrections are defined in the metal-poor regime. Similarly, the rMS is used to calibrate models at an intermediate metallicity between the bMS and the solar-metallicity cluster, M67. NGC~6791 is used as a fiducial case at a super-solar metallicity. These fiducial sequences are collectively used to model the metallicity dependence of the color corrections.

The cluster fiducial sequences in \citet{an:08} and \citet{bernard:14} are taken in the SDSS and PS1 systems, respectively. To employ the improved photometric calibration for the cluster photometry, SDSS DR6 photometry in each cluster's flanking fields is compared to DR14 photometry that is based on the `uber'-calibration \citep{finkbeiner:16}, and the mean differences are applied to the original sequences in \citet{an:08} to be consistent with the DR14 scale. The size of the corrections amount to a few hundredths of magnitudes. The adopted distance, metallicity, and reddening for the globular cluster sample (M15, M92, M13, M3, and M5, in increasing order of [Fe/H]) are the same as in \citet{an:09b,an:13}, and those for the open clusters (M67 and NGC~6791) are taken from \citet{an:19b}. For globular clusters, uniform errors in the color differences ($0.02$--$0.04$~mag) are adopted in each color index, which are propagated through errors in the adopted cluster parameters \citep[see][for a detailed error analysis]{an:09b}. The errors for the open cluster samples are those propagated from errors in \citet{an:19b}.

The advantage of {\it Gaia}'s double sequences over the cluster fiducial sequences is that they cover a large dynamic range of stellar mass based on an accurate distance scale. However, the mean metallicities of the bMS and rMS are not as precisely known as those of the calibrating cluster systems. Given that the observed colors depend on metallicity by $\Delta (\gr) / \Delta {\rm [Fe/H]} \sim 0.2$~mag~dex$^{-1}$, the metallicity of a calibrating system should be determined to a precision of $\sim0.1$~dex, in order to make calibrated isochrones useful. By cross-matching sources with SDSS/APOGEE \citep{majewski:17}, \citet{gaiahrd} found $\langle {\rm [Fe/H]} \rangle \sim -1.3$ and $\langle {\rm [Fe/H]} \rangle \sim -0.5$ for the bMS and rMS, respectively. \citet{sahlholdt:19} used photometric metallicities of red-giant stars, based on SkyMapper \citep{wolf:18} DR1, and found slightly lower metallicities: $\langle {\rm [Fe/H]} \rangle \sim -1.4$ for the bMS and $\langle {\rm [Fe/H]} \rangle \sim -0.7$ for the rMS.

\begin{figure}[hbt!] \centering \includegraphics[scale=0.56]{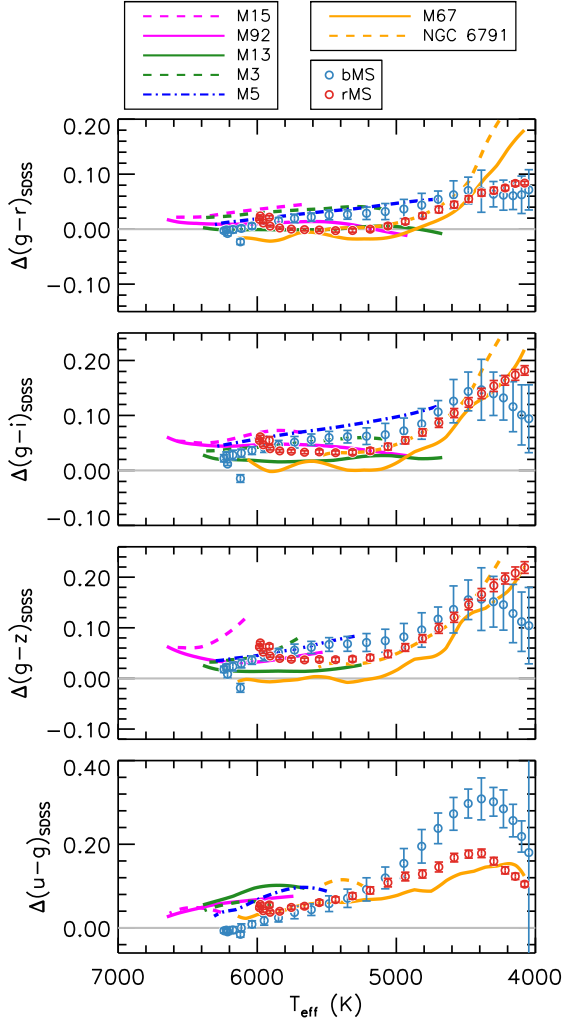} \caption{Color differences between observed fiducial lines and YREC+MARCS theoretical models (observation minus model). Dotted and solid lines are those derived from cluster fiducial sequences over a wide range of metallicity. The open circles with error bars show color differences of {\it Gaia}'s double sequences (see Figures~\ref{fig:gd}--\ref{fig:gd2}) .} \label{fig:cteff} \end{figure}

\begin{figure}[t!] \centering \includegraphics[scale=0.65]{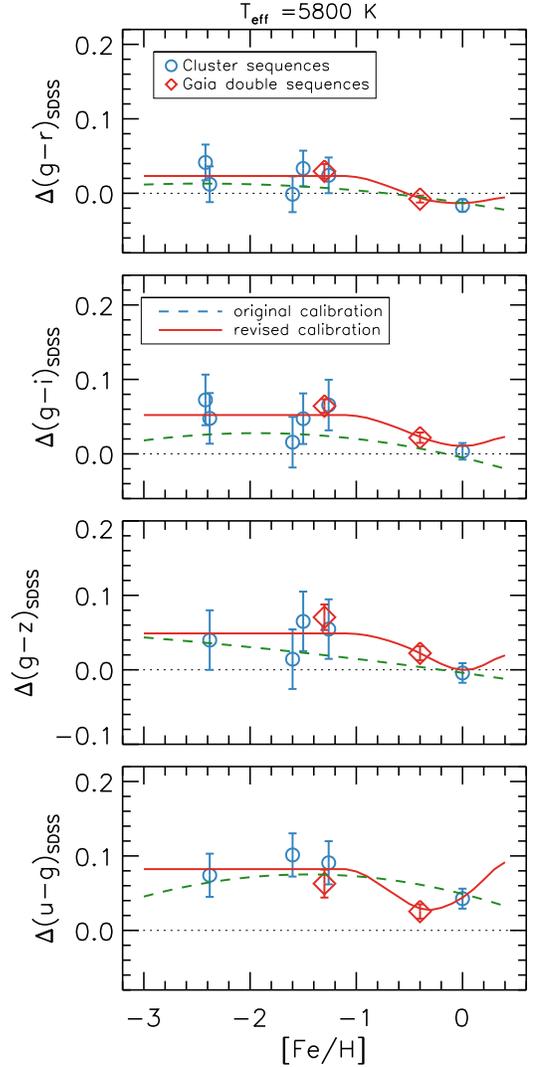} \caption{Derivation of empirical color corrections in the SDSS photometric system.  The blue open circles are color differences obtained from the key globular and open clusters. The red diamond symbols are those from {\it Gaia}'s double sequences. The adopted empirical correction function is shown by a red solid line. For comparison, the original correction is displayed by the green dashed line. The case at $\teff=5800$~K is shown in this example.} \label{fig:empcorr} \end{figure}

As described below, metallicities of the bMS and rMS are searched by minimizing changes of the color difference with respect to metallicity, and are adopted throughout this study. In Figure~\ref{fig:cteff}, the color offsets from the five globular clusters and two open clusters are shown over the entire $\teff$ range. Figure~\ref{fig:empcorr} shows mean color offsets at a fixed $\teff$ ($5800$~K in this example; see Appendix for more cases) as a function of [Fe/H]. For the globular cluster samples, there is no strong trend in the color residual over [Fe/H] at a given $\teff$. Since the bMS is on the metal-rich end of the globular clusters in our sample (more or less similar to that of M5), this leads to a plausible assumption that the bMS should have a similar color offset from the models as other globular clusters. The most satisfying color offsets in all color indices are found at [Fe/H] $=-1.3$ for the bMS. If a higher (lower) metallicity is adopted by $0.1$~dex, the bMS becomes too blue (red) with respect to the color residuals from the globular clusters. We adopt an age of $13$~Gyr for the bMS, based on an eyeball fit to the MS turn-off. The same age is adopted for other globular clusters, except M5 (12~Gyr).

On the other hand, the observed trend of the metal-rich systems is more dramatic, with a strong curvature in the color-$\teff$ relations for cool stars. The metallicity of the rMS, which is found between those of globular clusters (and the bMS) and M67, is determined by exploring different [Fe/H], while inspecting the resulting mean color offsets.  Because colors in the theoretical models are smoothly varying with metallicity, second-order terms (i.e., color corrections) should not exceed the underlying color changes of the models. If a higher (lower) [Fe/H] is taken, the rMS becomes too blue (or red) with respect to M67 (as well as the globular clusters), resulting in a strong curvature in the color corrections, and therefore model colors. If the metallicity is set to [Fe/H] $=-0.4$, as shown in Figure~\ref{fig:empcorr}, a smooth transition from M67 to globular clusters can be achieved. An age of 12~Gyr is found for the rMS, based on fitting of the original isochrone models to its MS turn-off.

The red line in Figure~\ref{fig:empcorr} shows an empirical correction function in each color index. As discussed above, we adopt a constant color correction at [Fe/H] $<-1$, and make a smooth transition to the color corrections from the rMS, M67, and NGC~6791 using a quadratic interpolation. The correction functions are computed at $\Delta \teff = 100$~K intervals. The impact of the Bayesian parallaxes in \citet{bailerjones:18} is negligible in the color calibration, as the color offsets change little for both the bMS ($0.01$--$0.02$~mag) and rMS ($<0.01$~mag), when Bayesian parallaxes are employed. These differences are included in the error bars for the double sequences in Figure~\ref{fig:empcorr}.

\begin{figure} \centering \includegraphics[scale=0.65]{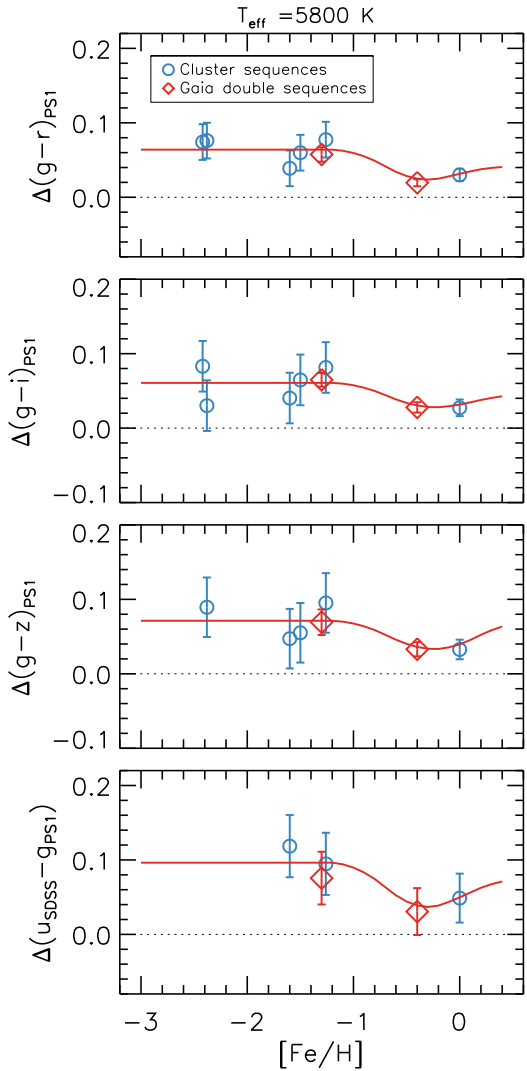} \caption{Same as in Figure~\ref{fig:empcorr}, but in the PS1 photometric system.} \label{fig:empcorrps} \end{figure}

The above procedure is repeated for the PS1 system (with SDSS $u$-band photometry), and the resulting color-correction functions are shown in Figure~\ref{fig:empcorrps} at $\teff=5800$~K. The overall trend of the color correction is quite similar to the case for SDSS, implying that the observed systematic departures from the models are unlikely caused by systematic errors in the photometry and/or filter-response functions; rather, they may be a sign of a scale error in $\teff$ and/or incorrect line-blanketing in the models.

\begin{figure} \centering \includegraphics[scale=0.46]{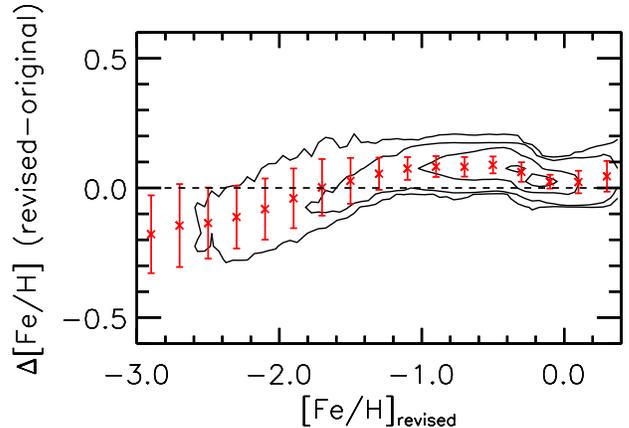} \caption{Comparison of photometric metallicities from $ugriz$, based on the revised calibration in this study with those from Paper~I.  All of the stars in the SDSS data set are included in the above comparison. The crosses and error bars show mean differences and standard deviations in bins of $0.2$~dex in [Fe/H]. Contour levels include 20\%, 60\%, 85\%, and 95\% of the sample.} \label{fig:comp} \end{figure}

Compared to the old calibration (green dashed line in Figure~\ref{fig:empcorr}), the revised calibration for metal-poor ([Fe/H] $<-1$) stars makes isochrones redder in $\gr$, $\gi$, and $\gz$, because mean color offsets from all globular cluster samples are taken in this study, instead of taking M92 as a baseline. The sense of this change is that the photometrically derived $\teff$ becomes higher at a given color. A higher $\teff$ can further lead to a longer distance and a higher photometric metallicity. The newer calibration, which is designed to directly follow M67 and NGC~6791, also has a sharper dip at $-1 \la {\rm [Fe/H]} \la 0$ in all color indices.

The net difference in metallicity between the original and the revised calibration is shown in Figure~\ref{fig:comp}. The mean differences in [Fe/H] are $\sim0.1$~dex at intermediate metallicities ($-1.4 \la {\rm [Fe/H]} \la -0.2$), and are $\sim0.2$~dex on the metal-poor side.  However, it is not trivial to determine the impact of the revised calibration, because of the interplay of the color corrections in the derivation of $\teff$, [Fe/H], and distance. Therefore, we construct stellar distributions in the [Fe/H] versus $v_\phi$ plane in the following section, in order to judge the consequence of the revised calibration. 

\clearpage

\section{Chemo-dynamical Distribution of the Galactic Halo}\label{sec:blueprint}

Based on the revised set of isochrones, metallicities and distances of individual stars are computed, employing the same procedure as in Paper~I. {\it Gaia}'s proper-motion data are used along the Galactic prime meridian, in combination with photometric distances, to estimate the rotational velocities of stars. The following analysis is restricted to $4$--$6$~kpc from the Galactic plane, to verify the accuracy of the current models, where a number of stellar populations in the Milky Way can reveal themselves on the [Fe/H]-$v_\phi$ plane with a proper normalization scheme (Paper~I), or after a masking of dominant stellar components (this study).

\begin{figure*}[hbt!] \centering \includegraphics[scale=0.32]{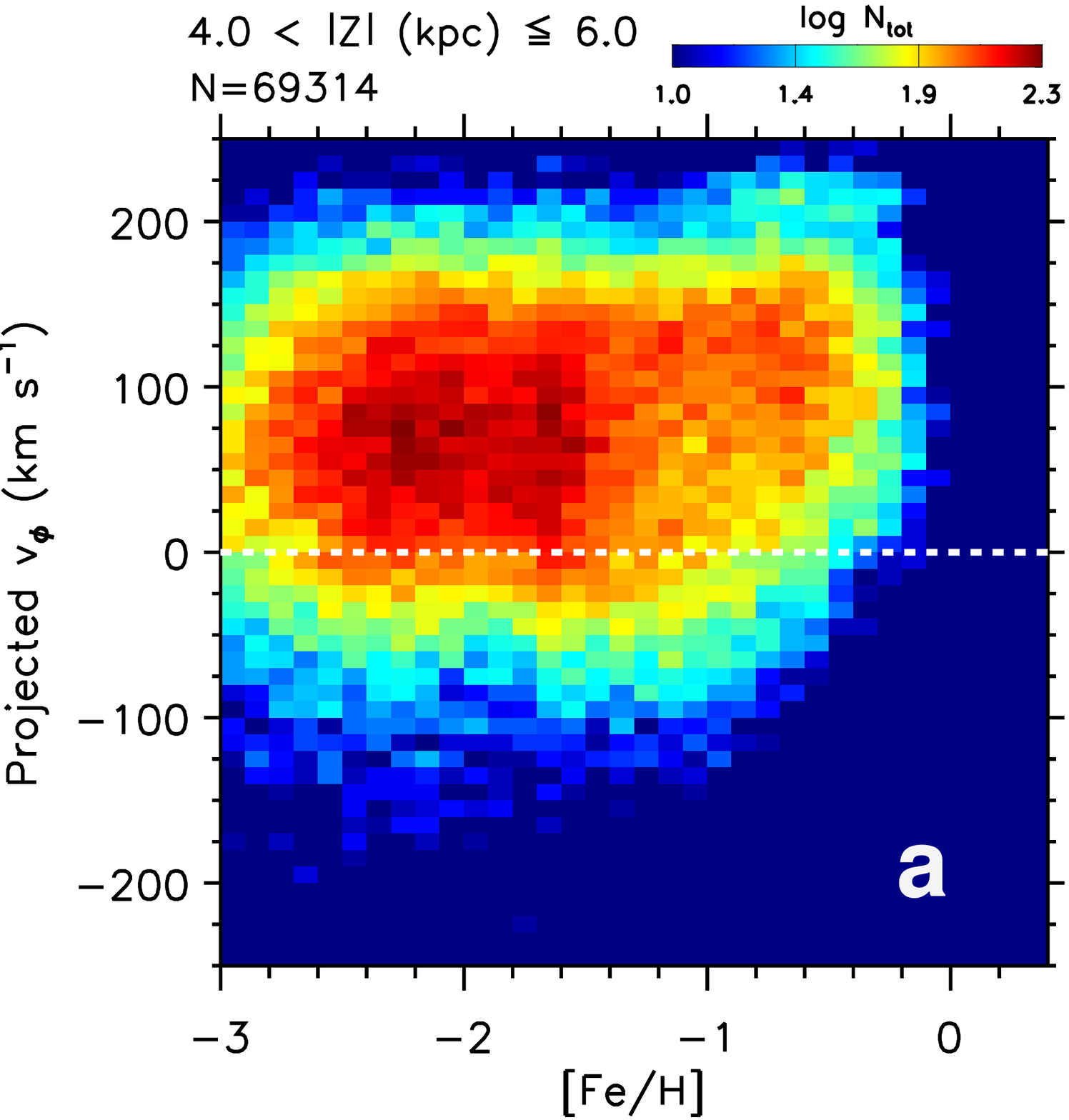} \includegraphics[scale=0.32]{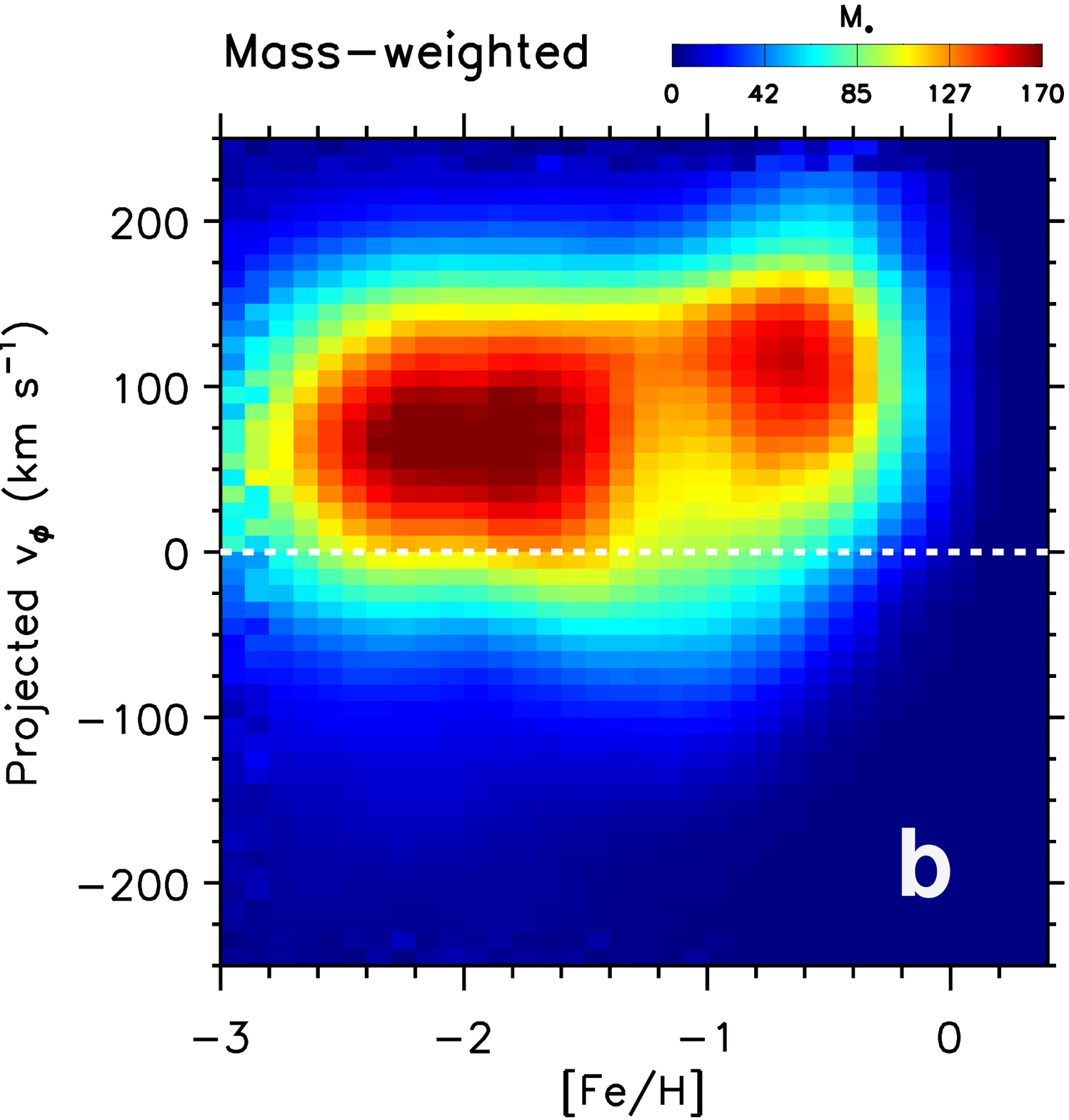} \includegraphics[scale=0.32]{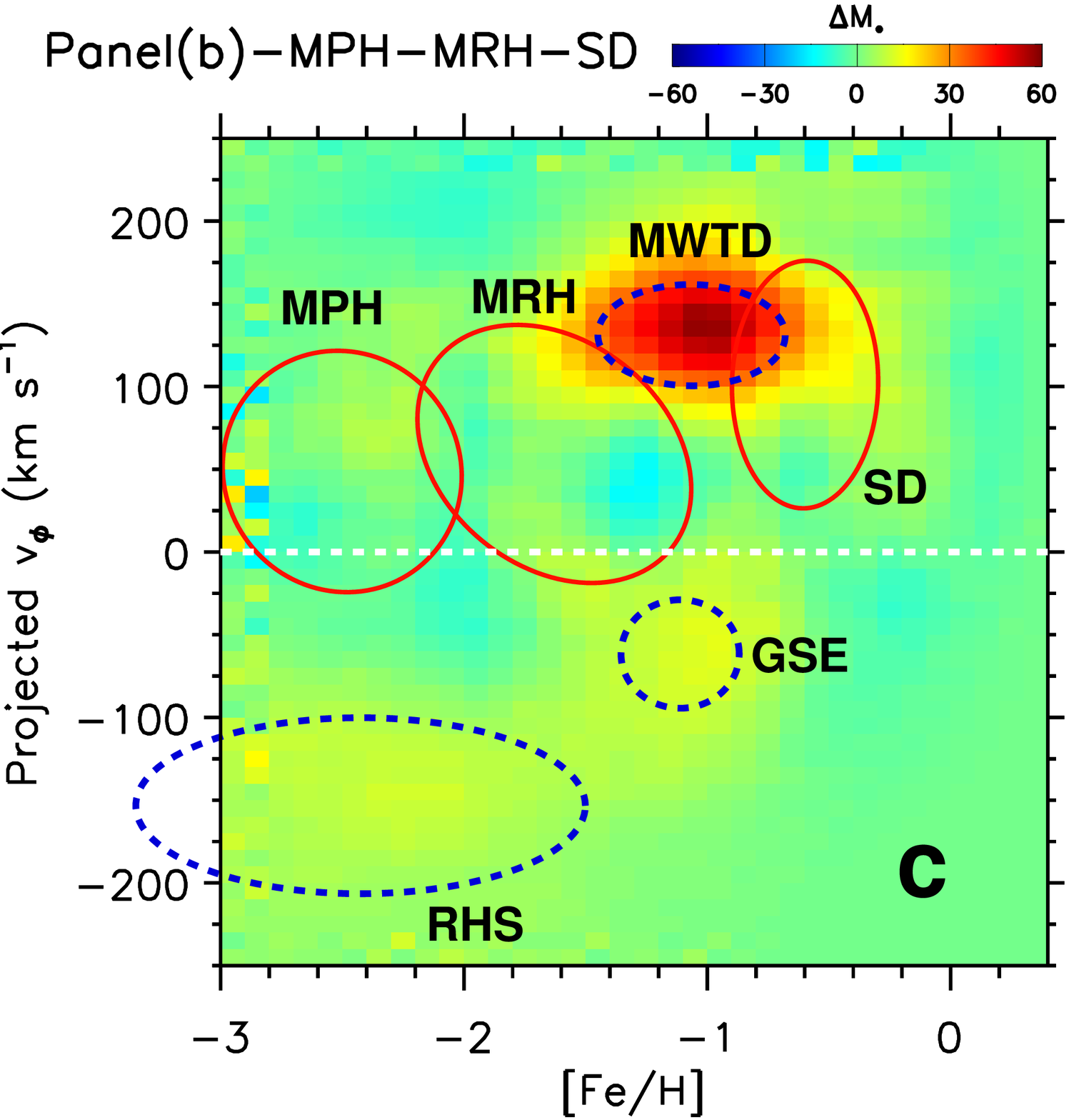} \includegraphics[scale=0.32]{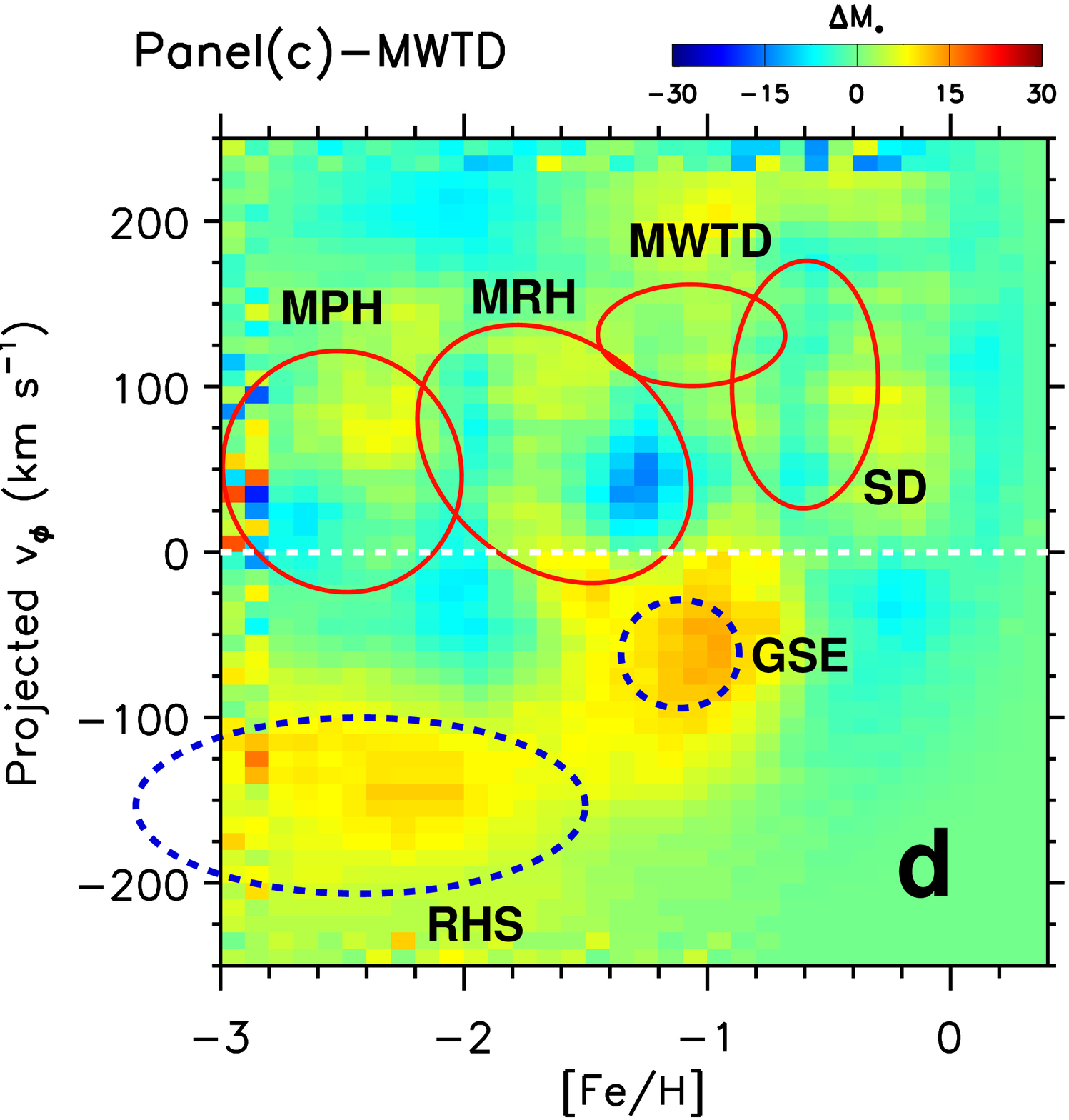} \includegraphics[scale=0.32]{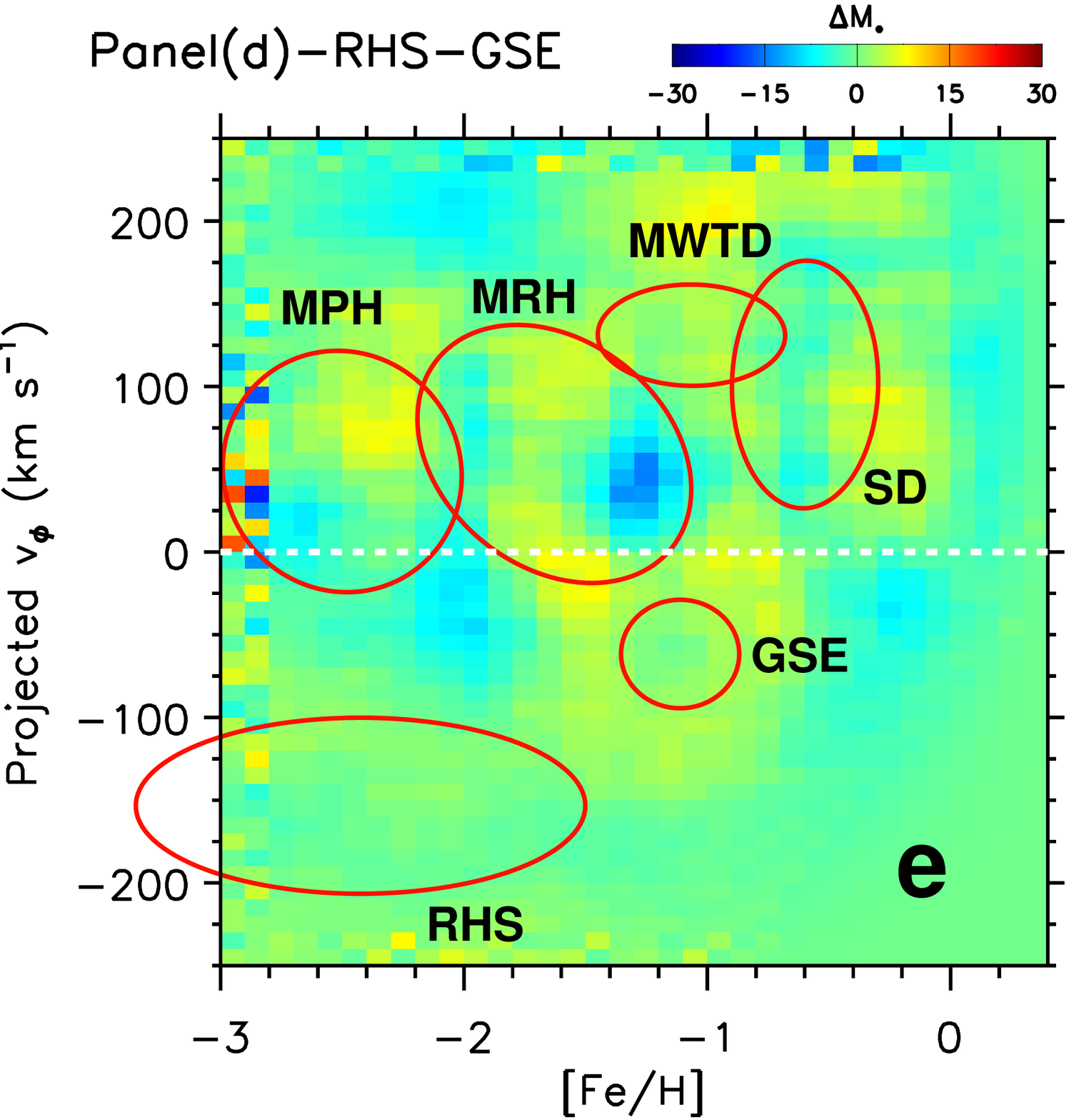} \includegraphics[scale=0.32]{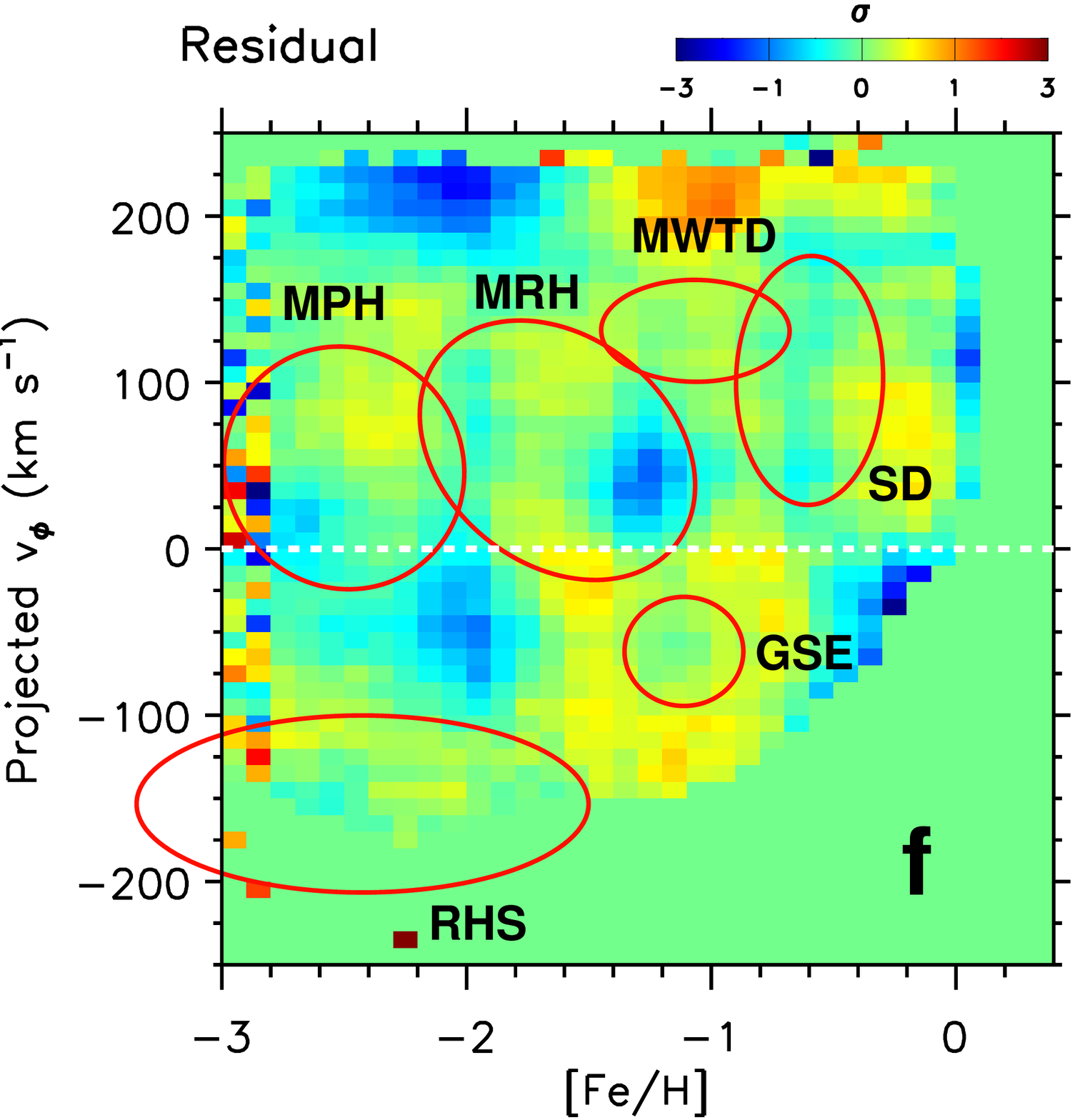} \caption{Decomposition of stellar populations in the Milky Way at $4$--$6$~kpc from the Galactic plane. Rotational velocities in the Galactocentric cylindrical coordinate system are computed using {\it Gaia}'s proper motions and photometric distances along the Galactic prime meridian.  Photometric distances and metallicities are derived using SDSS photometry. (a) A logarithmic number-density distribution of the stars.  (b) Same as in (a), but showing a linear density distribution, after correcting for bias in stellar mass (see text). (c) A number-density distribution after subtracting three major components -- the prograde metal-poor halo (MPH), the metal-rich halo (MRH), and the SD -- modeled using Gaussian distributions (red solid ellipses). The metal-weak thick disk (MWTD) is seen in this residual distribution, which is fit using a Gaussian ellipse (blue dotted ellipse). (d) A residual distribution after subtracting the MWTD in (c). The remaining stars are fit using two Gaussian functions (red dotted ellipses), each of which represents GSE and the retrograde halo structure(s) (RHS), respectively. (e) A residual distribution after subtracting all sub-components (solid ellipses) in (c) and (d). (f) The significance of the residual map in (e). Note that the color scales are different in all panels.} \label{fig:decomp} \end{figure*}

Figure~\ref{fig:decomp} shows the observed distribution of stars in the [Fe/H]-$v_\phi$ plane extracted using the revised isochrones, along with its decomposition into sub-populations. Panel~(a) shows a raw, unweighted stellar distribution of stars at $4 < |Z| \leq 6$~kpc on a logarithmic scale. In our sample, there is a luminosity bias for MS stars, according to which the observed metal-rich MS stars tend to be more massive than their metal-poor counterparts in a given distance bin.  As more-massive MS stars are outnumbered by less-massive stars in a standard mass function of a stellar population, this implies an under-estimation of the number of metal-rich stars in our sample. This problem is mitigated by multiplying the raw star count by $(M_*/0.7\ M_\odot)^{2.35}$, based on the \citet{salpeter:55} mass function. Here, $M_*$ is the median value of stellar mass in each bin, which usually span a narrow range of mass ($<0.1\ M_\odot$) at a given metallicity.  The scaled distribution based on this simple correction is shown in panel~(b). The impact of the bias correction is minimal, however, because the mass difference between the metal-poor and metal-rich star samples is small ($< 0.1\ M_\odot$).

In an attempt to characterize the observed distribution and isolate individual Galactic subpopulations, we fit mixtures of elliptical Gaussians to the [Fe/H]-$v_\phi$ plane. This analysis is done mostly for the purpose of demonstration; a more detailed modeling will be presented in the subsequent paper of this series (An et al., in preparation).

\begin{figure*}[hbt!] \centering \includegraphics[scale=0.32]{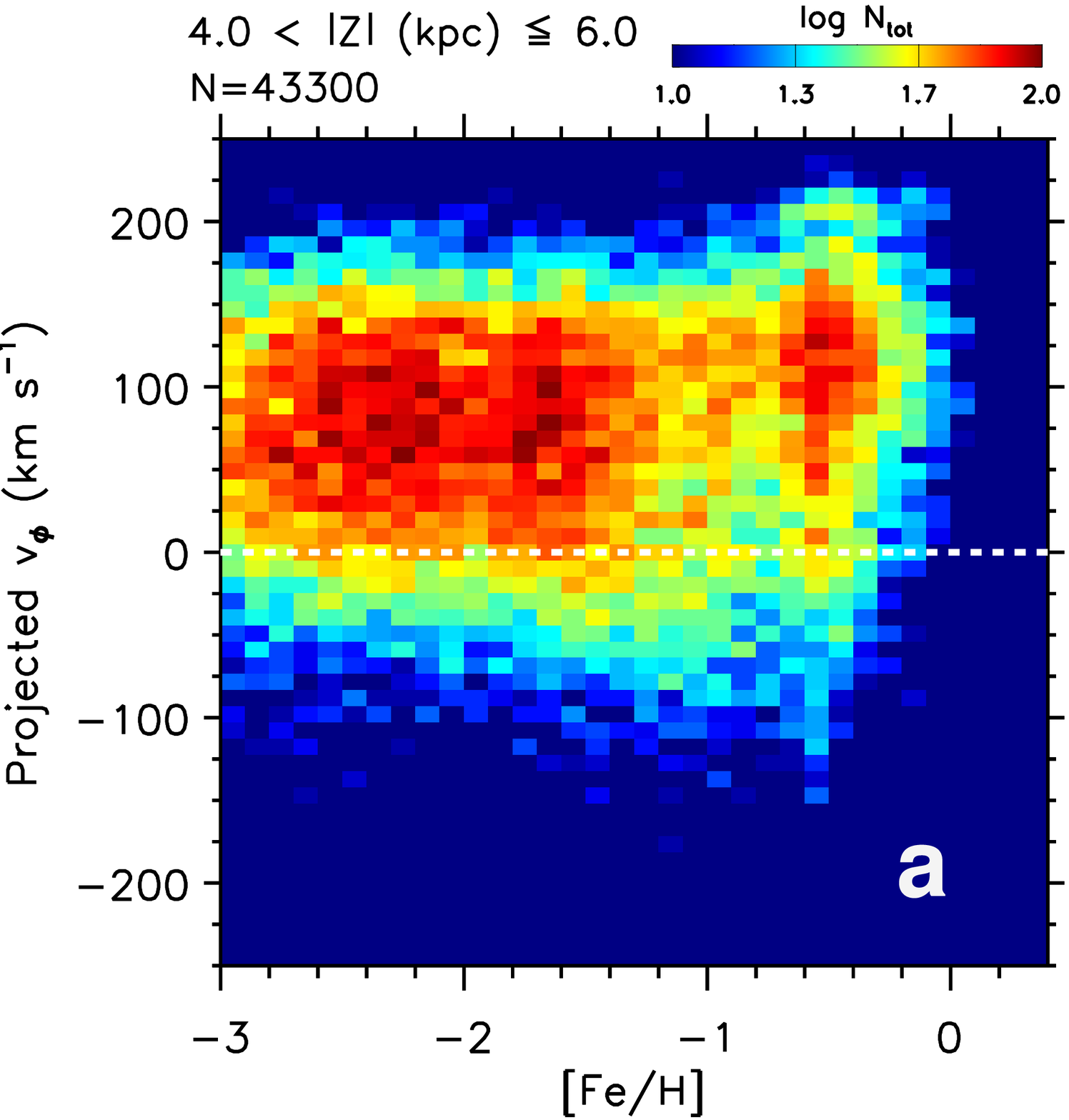} \includegraphics[scale=0.32]{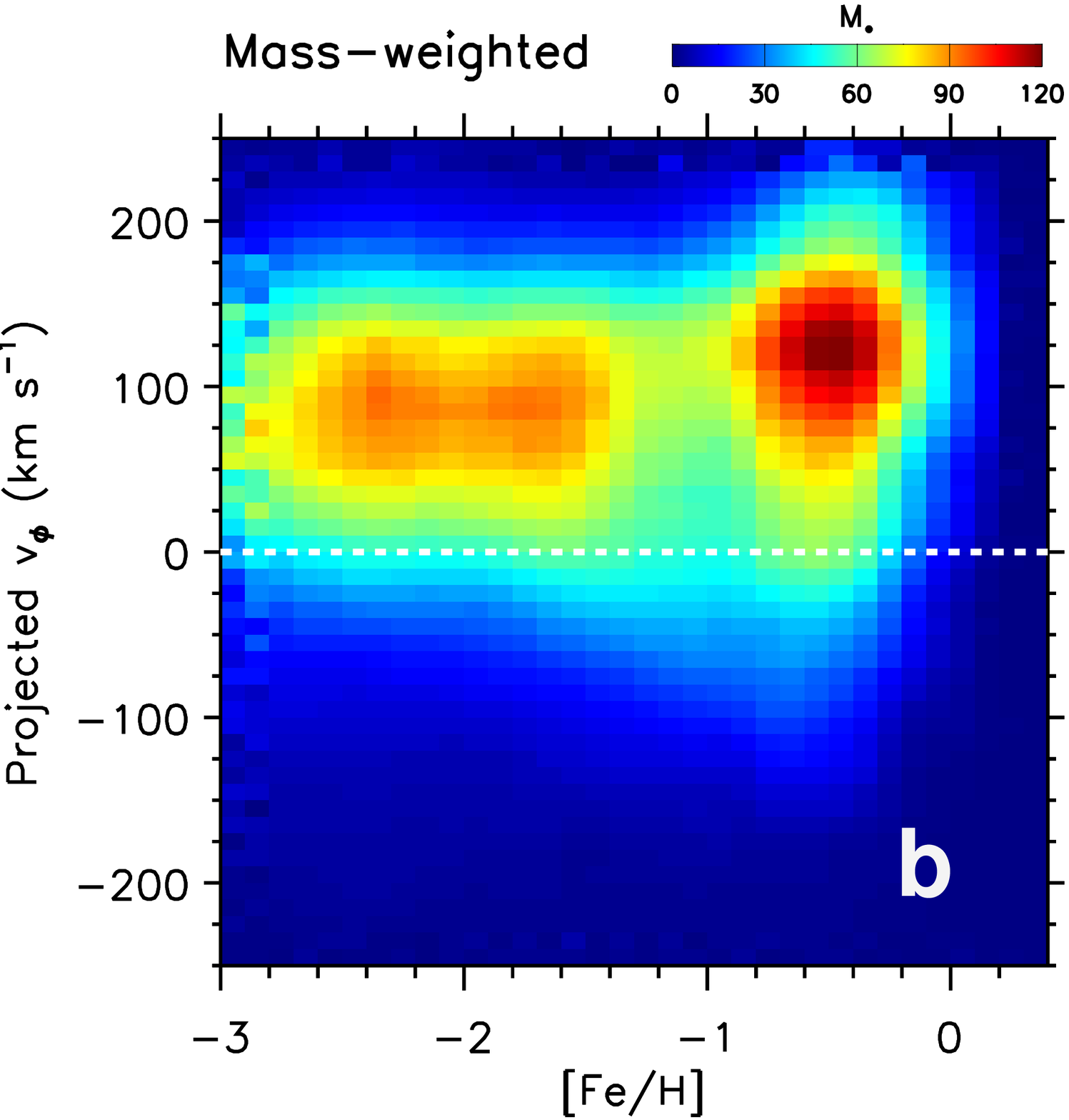} \includegraphics[scale=0.32]{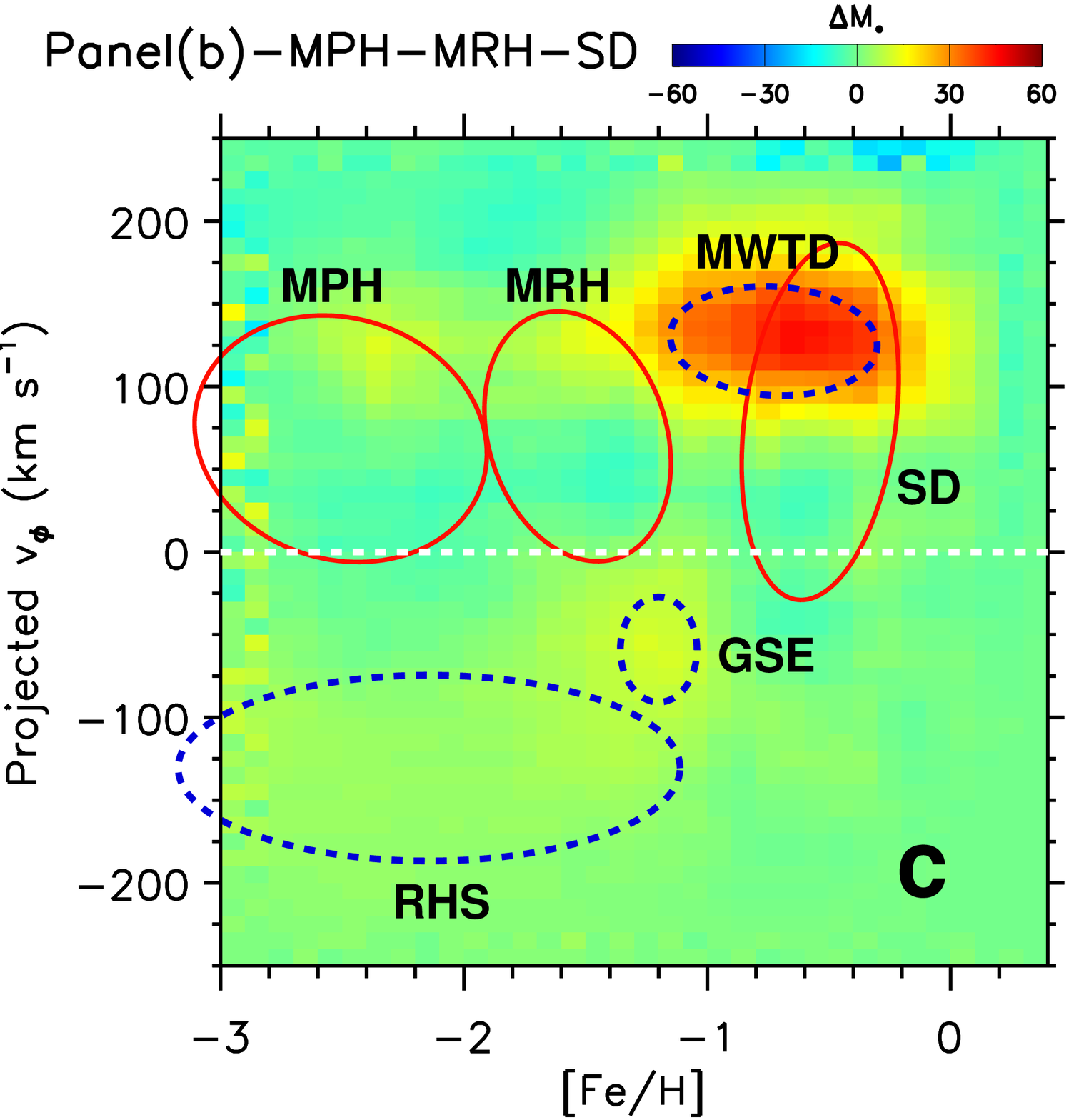} \includegraphics[scale=0.32]{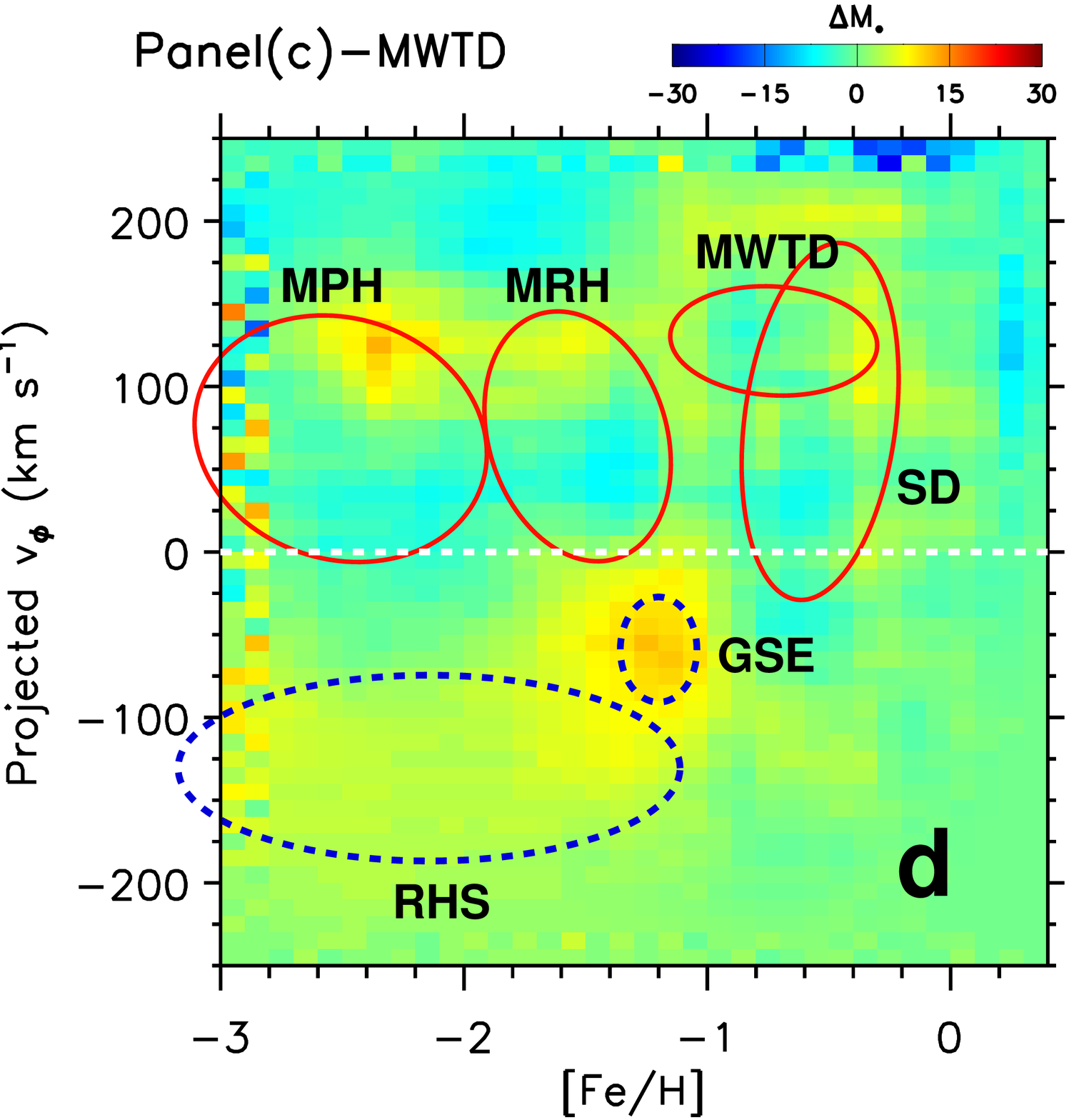} \includegraphics[scale=0.32]{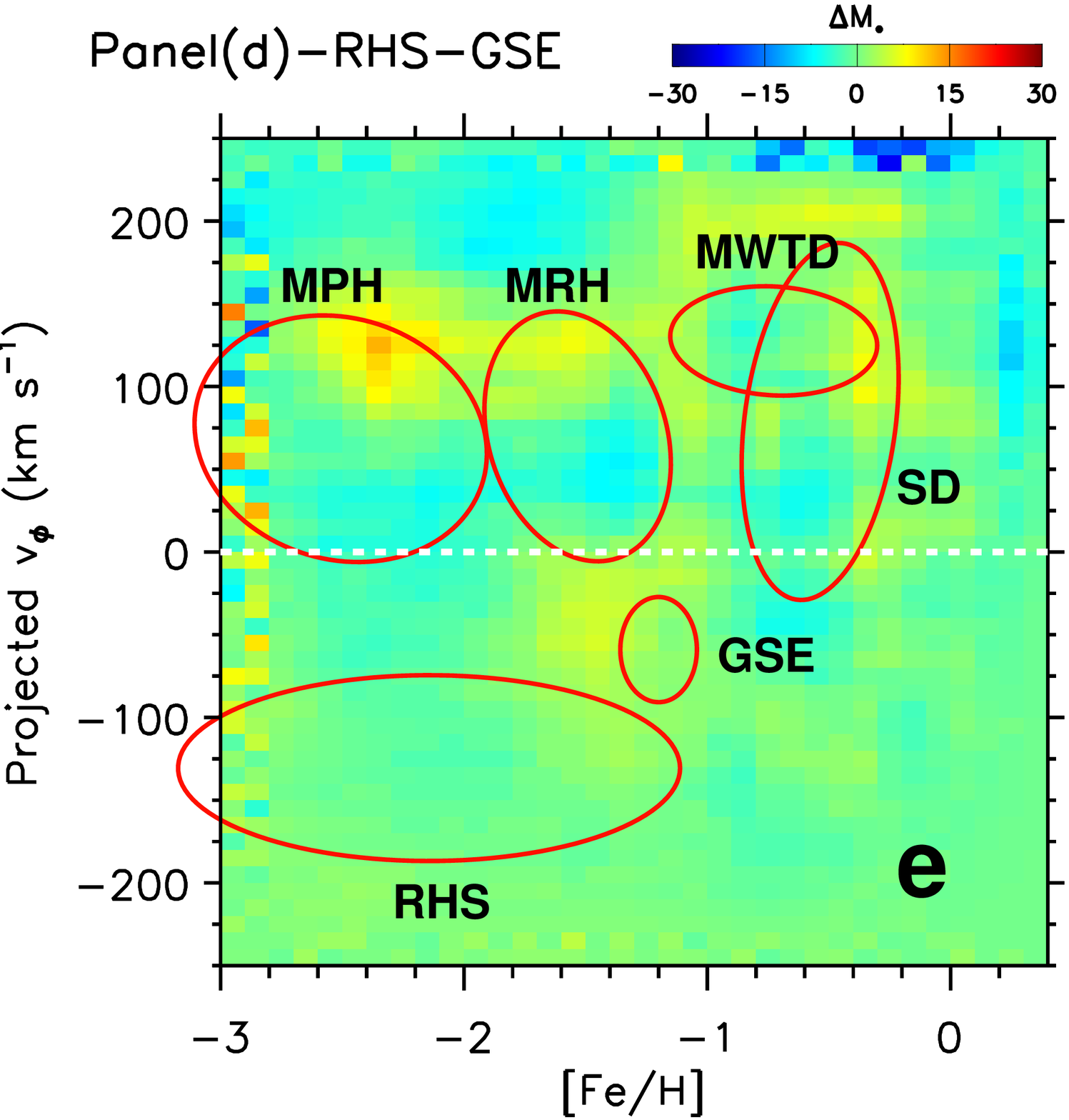} \includegraphics[scale=0.32]{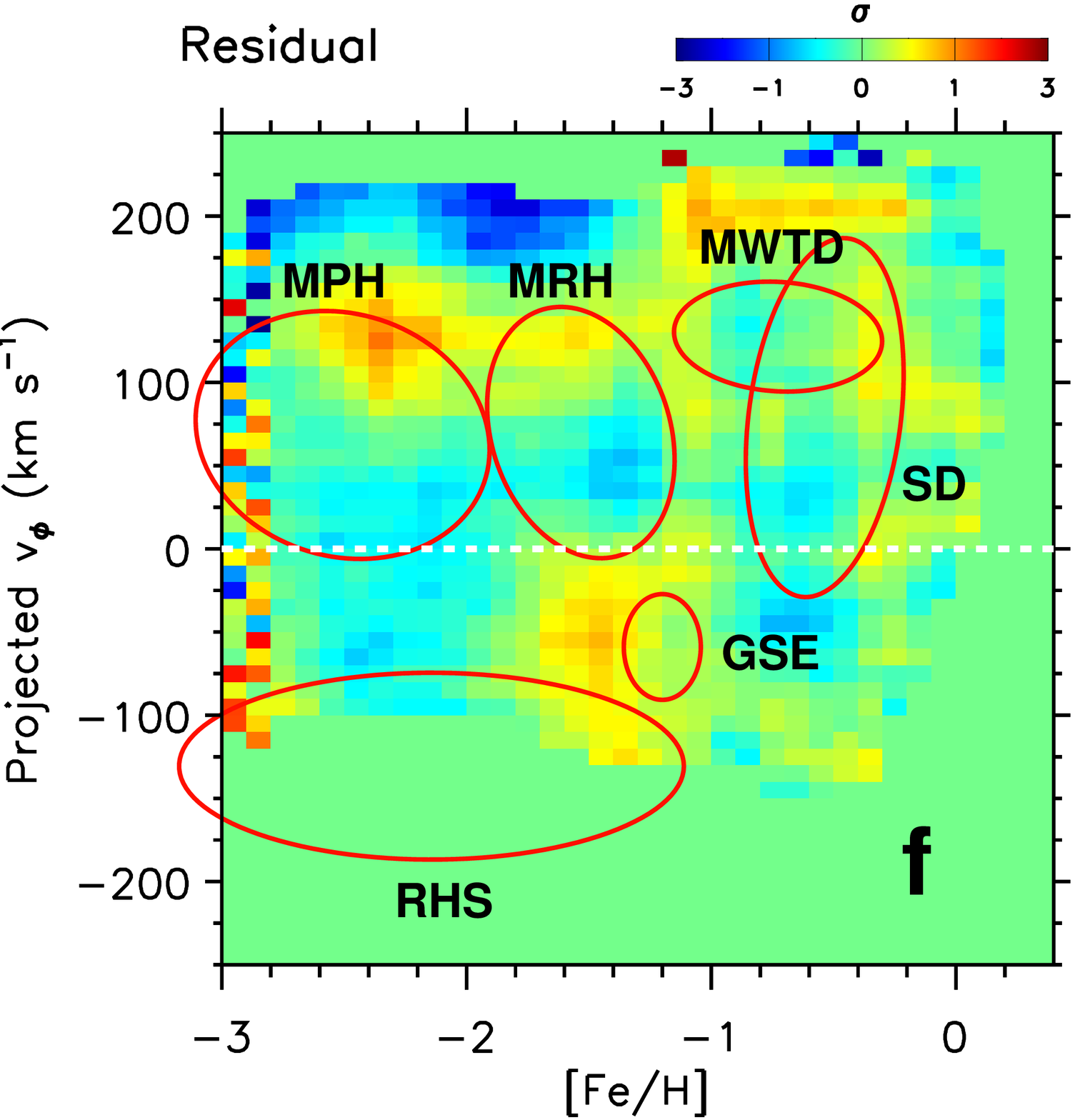} \caption{Same as in Figure~\ref{fig:decomp}, but based on the PS1 photometry.} \label{fig:decomp_ps1} \end{figure*}

We begin by modeling major structures visible in panel~(b): the metal-poor halo over a wide range of metallicity ($-2.6 < {\rm [Fe/H]} < -1.6$) and the metal-rich clump at [Fe/H] $\sim-0.6$, which is attributed to the SD according to Paper~I. They comprise approximately $90\%$ of the stars in this data set. The former exhibits double peaks at [Fe/H] $\sim-1.6$ and $-2.5$, reminiscent of the central values of the IH and OH populations in \citet{carollo:07,carollo:10}, respectively.  Indeed, a fit using a single Gaussian under-estimates the number of stars at both ends of the metallicity distribution, and at least two Gaussians are required to capture the observed distribution of the metal-poor stars. The solid red ellipses in panel~(c) show the positions and the extents of the three best-fitting Gaussians to the major structures -- the SD, the metal-rich, and the metal-poor halos on prograde orbits, respectively. The residual distribution after subtracting these three components is shown by the pixel colors. This step requires a few iterations, with fitting other components, including the metal-weak thick disk \citep[MWTD;][see below]{beers:14,carollo:19}.

Once these major structures are subtracted off, the residual distribution shown in panel~(c) is dominated by an elongated structure of stars with $-1.6 < {\rm [Fe/H]} < -0.8$ on prograde orbits ($v_\phi \sim 130$~km~s$^{-1}$). In Paper~I, the clump near the same $ v_\phi$ is attributed to the MWTD. The central position and the extent of the clump are somewhat sensitively affected by how the major components are fit by elliptical Gaussians. In this regard, it is reassuring that this volume of the parameter space is also occupied by stars associated with the MWTD from spectroscopic data \citep{carollo:19}.\footnote{This parameter space also overlaps with the Helmi streams \citep{helmi:99}, which is characterized by high vertical speeds with respect to the Galactic plane. However, the current data set, based on $|Z|-v_\phi-$[Fe/H], is not sufficient to discriminate such objects from the MWTD members.} The revised calibration makes the metallicity of MWTD slightly lower than in Paper~I, since the color-metallicity relations are adjusted by employing the rMS at the intermediate metallicity near the MWTD. The result of fitting an elliptical Gaussian is shown in panel~(c) by a blue dashed line, and the residual after subtraction is shown in panel~(d).

As shown in Panel~(d) of Figure~\ref{fig:decomp}, the retrograde halo substructure(s) and GSE begin to reveal themselves after removing the MWTD and narrowing down the range in the density plot. The blue dashed curves show the best-fitting elliptical Gaussians to these components, assuming that metal-poor halo stars in retrograde orbits can be modeled as a single population. The fitting residuals after subtracting these components are shown in panel~(e). The significance of the residual is computed assuming Poisson statistics, and is displayed in panel~(f). There are some weak signals left in panels~(e)--(f), some of which may be related to other substructures in the halo \citep[e.g., ][and references therein]{helmi:20,naidu:20}.

Figure~\ref{fig:decomp_ps1} shows the same [Fe/H]-$v_\phi$ distribution and its decomposition into sub-populations as in Figure~\ref{fig:decomp}, but based on the PS1 photometry and the newly calibrated set of isochrones. The locations and extents of the sub-populations from PS1 are quite similar to those from SDSS, indicating that the models in the PS1 $griz$ colors are accurately calibrated, although the same SDSS $u$-band data are used in both cases.  The two systems are at least on a self-consistent metallicity and distance scale with each other.

\input{tab3.tex}

Table~\ref{tab:tab3} shows the mean central positions and standard deviations of each component in [Fe/H] and $v_\phi$ from Figures~\ref{fig:decomp}--\ref{fig:decomp_ps1}. The errors indicate half of the differences. A fraction of each component is computed from a direct integration of the fitted profile. According to these estimates, about equal proportions of stars at $4 < |Z| < 6$~kpc belong to the SD, the metal-rich, and the metal-poor prograde halos, respectively. Only the remaining $10\%$ of stars can be found in the MWTD, GSE, and the retrograde structure(s). The fraction of GSE is small at this distance, because its main body primarily encompasses a more local volume.

\section{Summary and Discussion}\label{sec:summary}

\subsection{Empirically Calibrated Isochrones}

Previous and ongoing imaging surveys can provide useful information on the properties of stars to a significant depth, with a high volume-filling factor, and therefore can complement the relatively smaller spectroscopic survey data. Extracting fundamental stellar parameters from the photometry, however, requires a well-calibrated relation between a limited number of observables and physical quantities. As one of such approaches, our previous exercise relied on observations of well-studied Galactic star clusters to calibrate theoretically predicted quantities in stellar isochrones. This semi-empirical approach has been successfully applied to the SDSS data in Paper~I to map the distribution of halo stars on the metallicity versus rotational-velocity space, leading to clear identification of a number of distinct stellar populations.

Our previous work is ultimately tied to the {\it Hipparcos} \citep{esa:97} distance scale, since cluster distances are adopted from the subdwarf-fitting to globular clusters and the parallaxes of individual stars in open clusters. With the advent of {\it Gaia}, however, it becomes possible to utilize its exquisite parallaxes and other astrometric information in the isochrone calibration. This study provides the first look at the consequence of the revised calibration, which incorporates {\it Gaia}'s double sequences from stars with large rotational velocities.

{\it Gaia}'s double sequences open up a new possibility for sampling a single stellar population from a dynamical family of stars in the Milky Way, and utilizing it to define a fiducial color-magnitude relation. The observed double sequences are sufficiently narrow to define purely empirical color-magnitude relations in the native SDSS and PS1 photometric systems. In particular, the rMS covers a wide range of mass, down to $\sim0.3\ M_\odot$. Since these stars are found across a large survey area, the observed sequences are essentially free from small-scale fluctuations in the photometric and astrometric zero-point levels, which are the real advantages over using star clusters.

Our previous mapping result in Paper~I suggests that the bMS belongs to a combination of the IH and GSE, while the rMS is composed of the SD population. The metallicity difference between the two sequences is approximately $1$~dex in [Fe/H], providing a useful leverage on the metallicity dependence of stellar colors. When compared to theoretical stellar isochrones, we find that bMS consistently exhibits redder colors; the size of the color offset is consistent with those found from globular clusters. Because the metallicity of the bMS population is comparable to that of metal-rich globular clusters in the halo, and the colors change mildly in the metal-poor regime, the agreement confirms our adopted distance scale for the globular clusters and the accuracy of the cluster photometry.

In this work, we construct a set of isochrones in the SDSS and PS1 passbands with the revised color corrections, and use it to map the [Fe/H]-$v_\phi$ distribution of stars at $4$--$6$~kpc from the Galactic plane. Because a number of populations reveal themselves in this distance bin, we choose this sample of stars for the purpose of verifying the accuracy of the newly calibrated models. We find that both sets of isochrones in SDSS and PS1 (along with the SDSS $u$-band) produce qualitatively similar results to those in Paper~I. We also refine positions in the [Fe/H]-$v_\phi$ plane, and estimate the rough contributions of each population. Among various populations, the position and the extent of the MWTD are particularly well-captured in the revised map, because colors in the intermediate-metallicity range are better handled by the metal-rich sequence. This result strengthens the conclusion reached in Paper~I, that the MWTD is a separate component from the canonical TD \citep[see also][]{carollo:19}.

The photometric metallicity scale may differ from the spectroscopic metallicity scale. Unfortunately, there are a limited number of metal-poor MS stars with both homogeneous, high-resolution spectroscopic abundance analyses and good $ugriz$ photometry in the native filter systems. For relatively metal-rich stars ([Fe/H] $> -1$), the comparison with the GALAH survey \citep{buder:18} indicates that the systematic difference does not exceed $0.1$~dex (see also Figure~2 in Paper~I). The scale error may be as large as $0.1$--$0.3$~dex in the lower metallicities, judged from the comparisons to medium-resolution spectroscopic data from SDSS/SEGUE \citep{yanny:09} and high-resolution follow-up observations (C.\ M.\ Rockosi et~al., in preparation).  Inhomogeneous abundance ratios of Milky Way stars can also affect the photometric metallicity scale, since enhancement of $\alpha$-element abundances has a net effect of increasing the overall metallicity of a star \citep[e.g.,][]{kim:02}. Therefore, distinct [Fe/H]-[$\alpha$/Fe] sequences discovered in the Milky Way \citep[e.g.,][]{nissen:10, hayden:15,matsuno:19} can exhibit shifted [Fe/H] values in our mapping, unless specific [$\alpha$/Fe] are taken in the isochrone calibration.

Based on comparisons to SDSS/SEGUE spectroscopic metallicities, it can be seen that the precision of the models in the PS1 system is lower than the calibrated models in the SDSS $ugriz$ bands. The comparison of the PS1-based photometric metallicity with the spectroscopic data indicates that standard deviations of the metallicity difference ($\Delta {\rm [Fe/H]}=0.6$, $0.4$, and $0.2$~dex at [Fe/H]=$-2$, $-1$, and $0$, respectively) are $10\%$--$15\%$ larger for metal-poor stars ($-2 \la {\rm [Fe/H]} \la -1$), and are $50\%$ larger at solar metallicities than the case with the revised calibration in the SDSS $ugriz$. This mild deterioration in the models may originate from inaccurate cluster fiducial sequences and/or extinction corrections in the PS1 filter passbands, among other sources of possible systematic errors.

In the current data release of {\it Gaia} (DR2), good parallaxes are limited to relatively nearby stars, and do not provide statistically meaningful results for the majority of globular clusters, because of a non-negligible systematic error in the measurements \citep{gaiagc}. The color-$\teff$ relations can be further improved, once more accurate distances to the key open and globular clusters become available. Deep crowded-field photometry of the globular clusters will also be useful to accurately determine color-$\teff$ relations for low-mass metal-poor stars. On the contrary, the calibration will benefit from relatively `shallow' photometry of nearby clusters, because the upper MS in these well-studied clusters are often saturated in the broadband imaging surveys.

\subsection{Halo Dichotomy}

There is confusion in the literature on the relationship of the dual halos \citep{carollo:07,carollo:10,beers:12} to recently discovered structures in the halo. \citet{helmi:20} proposed a connection between the halo dichotomy and {\it Gaia}'s double sequences, according to which the IH is, by and large, the same component as the SD, and the OH is likely GSE. However, the dual halos were originally suggested based on a sample of metal-poor ([Fe/H] $ \la -1$) stars, which probably does not include the red-sequence stars. The peak metallicity of each component is also significantly different from each other ($\Delta {\rm [Fe/H]} \sim1$~dex), which is too large to be explained by a scale error in the spectroscopic metallicity measurements.

According to Figure~\ref{fig:decomp} (see also Paper~I), the metal-rich ($\langle {\rm [Fe/H]} \rangle\sim-1.6$) halo component in prograde rotation ($\langle v_\phi \rangle \sim60$~km\ s$^{-1}$) can plausibly be assigned to the IH. In addition, the original IH sample in \citet{carollo:07,carollo:10} may include a large number of stars in GSE, due to their similar metallicity ranges. Since GSE has a small scale height, the inclusion of the GSE would have led them to conclude that the IH has an oblate distribution, in contrast with a spherical distribution of the OH.

Meanwhile, the mean metallicities of the metal-poor halo component and the retrograde halo structure(s) are comparable to that of the OH ($\langle {\rm [Fe/H]} \rangle \sim-2.2$) in \citet{carollo:07, carollo:10}. According to our mapping, the former has a prograde rotation with $\langle v_\phi \rangle \sim +60$~km~s$^{-1}$, while the latter structures have $\langle v_\phi \rangle \sim -140$~km~s$^{-1}$.  Indeed, the asymmetric $v_\phi$ distribution of stars with [Fe/H] $<-2$ has previously been noted \citep{carollo:10,beers:12,an:13,an:15a}, according to which at least two Gaussian functions are required to model the observed $v_\phi$ distribution. \citet{carollo:07,carollo:10} argue that the OH has a net zero or a small retrograde rotation, which can be understood if the OH is composed of the prograde metal-poor halo and the retrograde structures, as our map shows. Recent re-analysis of the spectroscopic sample also supports this picture \citep{carollo:20}.

The retrograde structure(s) are likely related to other known components in the halo -- for example, Sequoia \citep{myeong:19} and/or Thamnos \citep{koppelman:19}. They may also include stars in the arc-like feature seen in the Toomre diagram \citep{helmi:18}, which is proposed to be a part of the tidal debris from GSE by these authors. In this regard, a long chain of stars connecting GSE and the retrograde structure(s) in panel~(d) of Figure~\ref{fig:decomp} may reflect a metallicity gradient of the progenitor(s) of GSE. On the other hand, there appears no matching substructure to the prograde metal-poor halo; although detailed dynamical analysis in \citet{limberg:20} reveals a number of cold stellar streams of metal-poor stars, their contribution may not be enough to account for the entire population.

In the following work of this series, we will continue our effort to constrain chemo-dynamical properties of the stellar populations in the local halo, by applying the revised set of empirically calibrated isochrones presented in this paper.

\acknowledgements

We thank Donald M.\ Terndrup for useful comments. D.A.\ thanks Coryn Bailer-Jones for kindly providing extensive data tables with Bayesian parallaxes. D.A.\ acknowledges support provided by the National Research Foundation of Korea to the Center for Galaxy Evolution Research (2017R1A5A1070354) and by Basic Science Research Program through the National Research Foundation of Korea (NRF) funded by the Ministry of Education (NRF-2018R1D1A1A02085433). T.C.B.\ acknowledges partial support from grant PHY 14-30152 (Physics Frontier Center/JINA-CEE), awarded by the U.S. National Science Foundation.

Funding for the Sloan Digital Sky Survey IV has been provided by the Alfred P. Sloan Foundation, the U.S. Department of Energy Office of Science, and the Participating Institutions. SDSS acknowledges support and resources from the Center for High-Performance Computing at the University of Utah. The SDSS web site is www.sdss.org.

SDSS is managed by the Astrophysical Research Consortium for the Participating Institutions of the SDSS Collaboration including the Brazilian Participation Group, the Carnegie Institution for Science, Carnegie Mellon University, Center for Astrophysics | Harvard \& Smithsonian (CfA), the Chilean Participation Group, the French Participation Group, Instituto de Astrofísica de Canarias, The Johns Hopkins University, Kavli Institute for the Physics and Mathematics of the Universe (IPMU) / University of Tokyo, the Korean Participation Group, Lawrence Berkeley National Laboratory, Leibniz Institut für Astrophysik Potsdam (AIP), Max-Planck-Institut für Astronomie (MPIA Heidelberg), Max-Planck-Institut für Astrophysik (MPA Garching), Max-Planck-Institut für Extraterrestrische Physik (MPE), National Astronomical Observatories of China, New Mexico State University, New York University, University of Notre Dame, Observatório Nacional / MCTI, The Ohio State University, Pennsylvania State University, Shanghai Astronomical Observatory, United Kingdom Participation Group, Universidad Nacional Autónoma de México, University of Arizona, University of Colorado Boulder, University of Oxford, University of Portsmouth, University of Utah, University of Virginia, University of Washington, University of Wisconsin, Vanderbilt University, and Yale University.

This work presents results from the European Space Agency (ESA) space mission Gaia. Gaia data are being processed by the Gaia Data Processing and Analysis Consortium (DPAC). Funding for the DPAC is provided by national institutions, in particular the institutions participating in the Gaia MultiLateral Agreement (MLA). The Gaia mission website is https://www.cosmos.esa.int/gaia. The Gaia archive website is https://archives.esac.esa.int/gaia.

The Pan-STARRS1 Surveys (PS1) and the PS1 public science archive have been made possible through contributions by the Institute for Astronomy, the University of Hawaii, the Pan-STARRS Project Office, the Max-Planck Society and its participating institutes, the Max Planck Institute for Astronomy, Heidelberg and the Max Planck Institute for Extraterrestrial Physics, Garching, The Johns Hopkins University, Durham University, the University of Edinburgh, the Queen's University Belfast, the Harvard-Smithsonian Center for Astrophysics, the Las Cumbres Observatory Global Telescope Network Incorporated, the National Central University of Taiwan, the Space Telescope Science Institute, the National Aeronautics and Space Administration under Grant No. NNX08AR22G issued through the Planetary Science Division of the NASA Science Mission Directorate, the National Science Foundation Grant No. AST-1238877, the University of Maryland, Eotvos Lorand University (ELTE), the Los Alamos National Laboratory, and the Gordon and Betty Moore Foundation.

\clearpage

\appendix

\begin{figure*}[t!] \centering \includegraphics[scale=0.65]{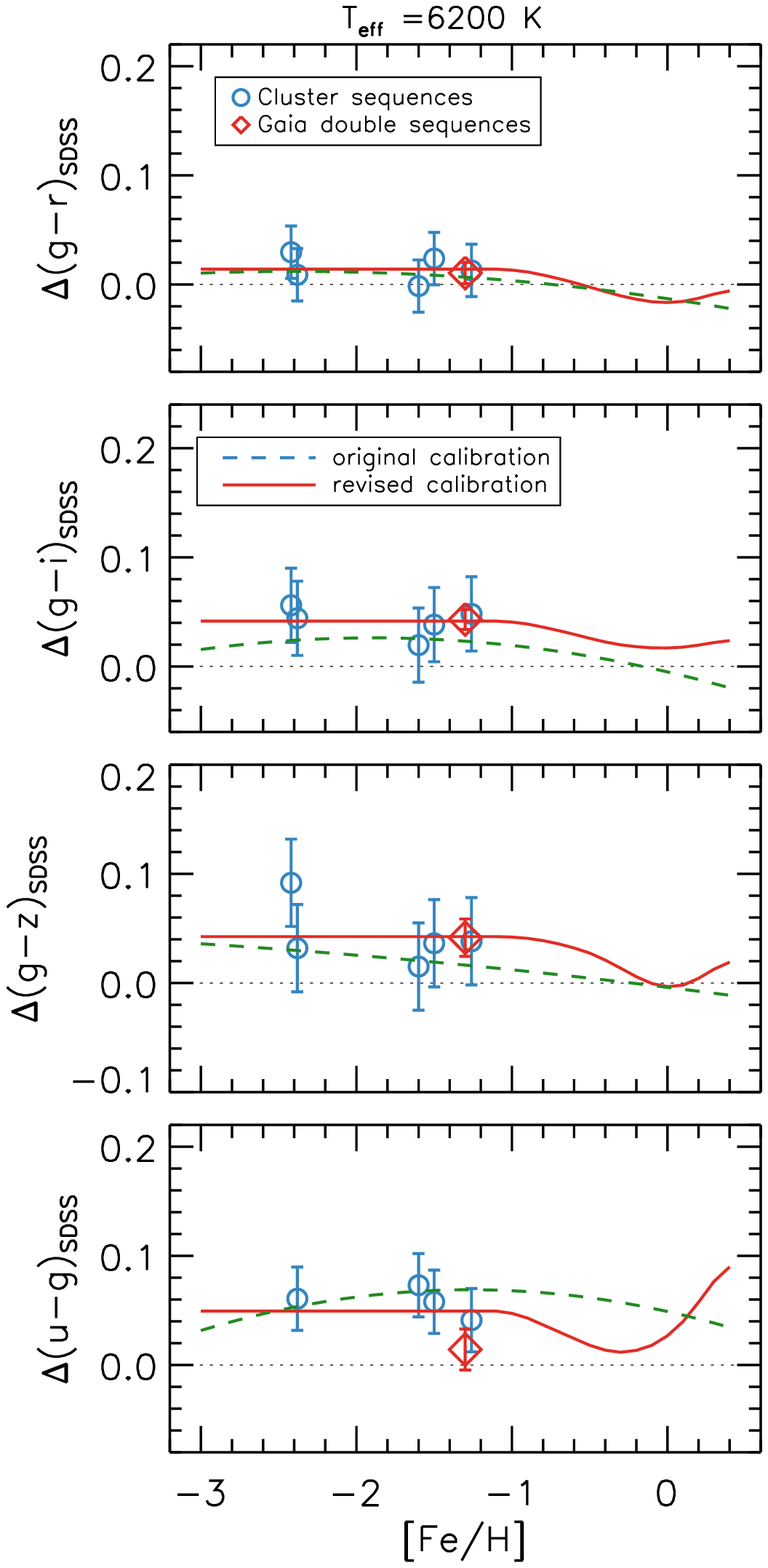} \includegraphics[scale=0.65]{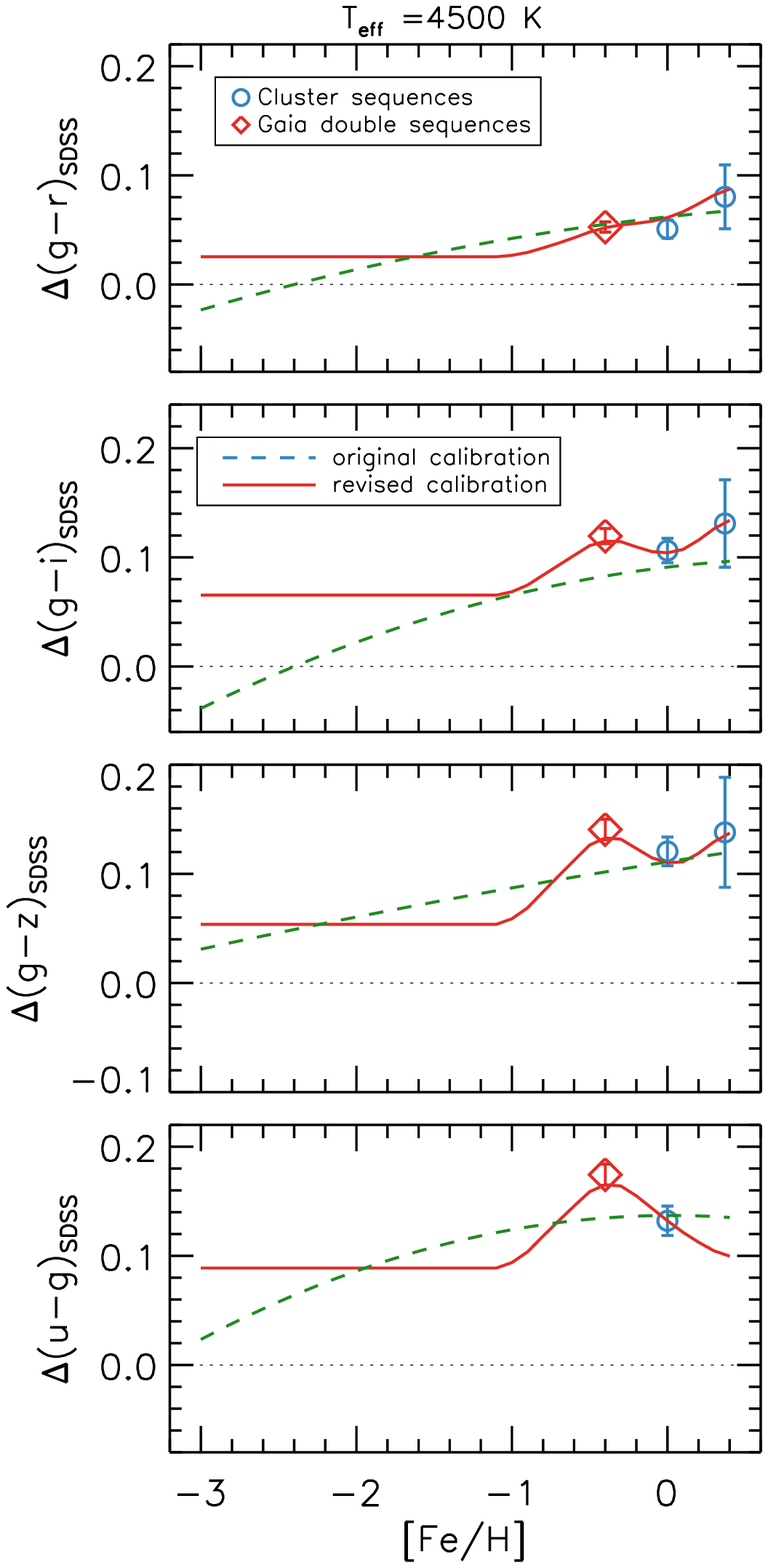} \caption{Same as in Figure~\ref{fig:empcorr}, but at $\teff=6200$~K (left) and $4500$~K (right).} \label{fig:empcorr2} \end{figure*}

In Figure~\ref{fig:empcorr2}, empirical color-correction functions other than the representative case in Figure~\ref{fig:empcorr} ($\teff=5800$~K) are shown at $\teff=6200$~K (left) and $4500$~K (right).

{}

\end{document}

%% file: tab1.tex
\begin{deluxetable*}{ccccccccc}
\tablecaption{{\it Gaia} Double Sequences in the SDSS System\label{tab:tab1}}
\tabletypesize{\scriptsize}
\tablehead{
   \colhead{} &
   \multicolumn{4}{c}{bMS} &
   \multicolumn{4}{c}{rMS} \\
   \cline{2-5}
   \cline{6-9}
   \colhead{$M_r$} &
   \colhead{$u\, -\, g$} & 
   \colhead{$g\, -\, r$} &
   \colhead{$g\, -\, i$} &
   \colhead{$g\, -\, z$} &
   \colhead{$u\, -\, g$} &
   \colhead{$g\, -\, r$} &
   \colhead{$g\, -\, i$} &
   \colhead{$g\, -\, z$}
}
\startdata
$3.300$&$0.955$&$0.290$&$0.380$&$0.419$&$1.248$&$0.443$&$0.580$&$0.632$ \\
$3.500$&$0.941$&$0.282$&$0.371$&$0.398$&$1.209$&$0.420$&$0.551$&$0.595$ \\
$3.700$&$0.929$&$0.275$&$0.364$&$0.382$&$1.174$&$0.401$&$0.527$&$0.563$ \\
$3.900$&$0.917$&$0.272$&$0.362$&$0.373$&$1.148$&$0.387$&$0.510$&$0.542$ \\
$4.100$&$0.908$&$0.273$&$0.366$&$0.375$&$1.133$&$0.381$&$0.502$&$0.532$ \\
$4.300$&$0.903$&$0.280$&$0.378$&$0.388$&$1.131$&$0.382$&$0.506$&$0.536$ \\
$4.500$&$0.902$&$0.292$&$0.397$&$0.410$&$1.143$&$0.392$&$0.520$&$0.553$ \\
$4.700$&$0.908$&$0.309$&$0.422$&$0.442$&$1.167$&$0.408$&$0.543$&$0.581$ \\
$4.900$&$0.922$&$0.331$&$0.453$&$0.481$&$1.206$&$0.429$&$0.575$&$0.620$ \\
$5.100$&$0.946$&$0.358$&$0.489$&$0.527$&$1.258$&$0.456$&$0.613$&$0.667$ \\
$5.300$&$0.981$&$0.388$&$0.531$&$0.579$&$1.324$&$0.488$&$0.658$&$0.722$ \\
$5.500$&$1.028$&$0.422$&$0.578$&$0.637$&$1.405$&$0.524$&$0.709$&$0.786$ \\
$5.700$&$1.089$&$0.458$&$0.630$&$0.700$&$1.498$&$0.567$&$0.768$&$0.857$ \\
$5.900$&$1.164$&$0.497$&$0.685$&$0.768$&$1.605$&$0.615$&$0.835$&$0.939$ \\
$6.100$&$1.254$&$0.539$&$0.744$&$0.841$&$1.724$&$0.672$&$0.913$&$1.032$ \\
$6.300$&$1.361$&$0.585$&$0.807$&$0.920$&$1.851$&$0.737$&$1.001$&$1.137$ \\
$6.500$&$1.482$&$0.637$&$0.878$&$1.007$&$1.981$&$0.808$&$1.100$&$1.255$ \\
$6.700$&$1.614$&$0.696$&$0.960$&$1.106$&$2.107$&$0.885$&$1.208$&$1.383$ \\
$6.900$&$1.752$&$0.763$&$1.055$&$1.218$&$2.220$&$0.963$&$1.320$&$1.517$ \\
$7.100$&$1.890$&$0.837$&$1.162$&$1.341$&$2.312$&$1.038$&$1.433$&$1.654$ \\
$7.300$&$2.021$&$0.913$&$1.276$&$1.472$&$2.380$&$1.107$&$1.541$&$1.787$ \\
$7.500$&$2.137$&$0.988$&$1.387$&$1.600$&$2.424$&$1.166$&$1.643$&$1.914$ \\
$7.700$&$2.232$&$1.055$&$1.486$&$1.717$&$2.447$&$1.215$&$1.735$&$2.032$ \\
$7.900$&$2.301$&$1.111$&$1.569$&$1.818$&$2.456$&$1.253$&$1.819$&$2.140$ \\
$8.100$&$2.345$&$1.159$&$1.635$&$1.900$&$2.458$&$1.283$&$1.895$&$2.241$ \\
$8.300$&$2.366$&$1.202$&$1.689$&$1.969$&$2.457$&$1.304$&$1.965$&$2.335$ \\
$8.500$&$2.373$&$1.243$&$1.736$&$2.030$&$2.456$&$1.320$&$2.030$&$2.424$ \\
$8.700$&$2.371$&$1.284$&$1.782$&$2.087$&$2.456$&$1.331$&$2.091$&$2.509$ \\
$8.900$&\nodata&\nodata&\nodata&\nodata&$2.455$&$1.338$&$2.149$&$2.592$ \\
$9.100$&\nodata&\nodata&\nodata&\nodata&$2.453$&$1.343$&$2.204$&$2.670$ \\
$9.300$&\nodata&\nodata&\nodata&\nodata&$2.450$&$1.347$&$2.254$&$2.744$ \\
$9.500$&\nodata&\nodata&\nodata&\nodata&$2.448$&$1.350$&$2.300$&$2.814$ \\
$9.700$&\nodata&\nodata&\nodata&\nodata&$2.447$&$1.353$&$2.344$&$2.880$ \\
$9.900$&\nodata&\nodata&\nodata&\nodata&$2.449$&$1.356$&$2.387$&$2.945$ \\
$10.100$&\nodata&\nodata&\nodata&\nodata&$2.453$&$1.359$&$2.430$&$3.010$ \\ 
$10.300$&\nodata&\nodata&\nodata&\nodata&$2.459$&$1.364$&$2.473$&$3.074$ \\ 
$10.500$&\nodata&\nodata&\nodata&\nodata&$2.465$&$1.369$&$2.517$&$3.138$ \\ 
$10.700$&\nodata&\nodata&\nodata&\nodata&$2.470$&$1.375$&$2.562$&$3.204$ \\ 
$10.900$&\nodata&\nodata&\nodata&\nodata&$2.473$&$1.380$&$2.607$&$3.270$ \\ 
$11.100$&\nodata&\nodata&\nodata&\nodata&$2.475$&$1.386$&$2.652$&$3.337$ \\ 
\enddata
\end{deluxetable*}

%% file: tab2.tex
\begin{deluxetable*}{ccccccccc}
\tablecaption{{\it Gaia} Double Sequences in the PS1 System\label{tab:tab2}}
\tabletypesize{\scriptsize}
\tablehead{
   \colhead{} &
   \multicolumn{4}{c}{bMS} &
   \multicolumn{4}{c}{rMS} \\
   \cline{2-5}
   \cline{6-9}
   \colhead{$M_r (PS1)$} &
   \colhead{$u_{SDSS}\, -\, g_{PS1}$} & 
   \colhead{$(g\, -\, r)_{PS1}$} &
   \colhead{$(g\, -\, i)_{PS1}$} &
   \colhead{$(g\, -\, z)_{PS1}$} &
   \colhead{$u_{SDSS}\, -\, g_{PS1}$} &
   \colhead{$(g\, -\, r)_{PS1}$} &
   \colhead{$(g\, -\, i)_{PS1}$} &
   \colhead{$(g\, -\, z)_{PS1}$}
}
\startdata
$3.300$&$0.970$&$0.265$&$0.330$&$0.330$&$1.290$&$0.389$&$0.526$&$0.545$ \\ 
$3.500$&$0.964$&$0.253$&$0.322$&$0.325$&$1.251$&$0.368$&$0.494$&$0.512$ \\ 
$3.700$&$0.958$&$0.245$&$0.317$&$0.322$&$1.217$&$0.349$&$0.467$&$0.484$ \\ 
$3.900$&$0.951$&$0.240$&$0.317$&$0.321$&$1.192$&$0.337$&$0.448$&$0.464$ \\ 
$4.100$&$0.944$&$0.242$&$0.323$&$0.326$&$1.177$&$0.331$&$0.440$&$0.455$ \\ 
$4.300$&$0.938$&$0.250$&$0.335$&$0.337$&$1.176$&$0.333$&$0.443$&$0.457$ \\ 
$4.500$&$0.936$&$0.264$&$0.353$&$0.355$&$1.188$&$0.341$&$0.456$&$0.472$ \\ 
$4.700$&$0.940$&$0.281$&$0.376$&$0.381$&$1.213$&$0.355$&$0.477$&$0.496$ \\ 
$4.900$&$0.954$&$0.301$&$0.403$&$0.413$&$1.253$&$0.373$&$0.505$&$0.529$ \\ 
$5.100$&$0.981$&$0.324$&$0.435$&$0.451$&$1.307$&$0.396$&$0.540$&$0.570$ \\ 
$5.300$&$1.020$&$0.349$&$0.472$&$0.495$&$1.377$&$0.424$&$0.581$&$0.618$ \\ 
$5.500$&$1.072$&$0.375$&$0.512$&$0.543$&$1.463$&$0.456$&$0.627$&$0.674$ \\ 
$5.700$&$1.137$&$0.403$&$0.555$&$0.596$&$1.563$&$0.493$&$0.681$&$0.737$ \\ 
$5.900$&$1.218$&$0.433$&$0.604$&$0.655$&$1.678$&$0.535$&$0.742$&$0.810$ \\ 
$6.100$&$1.316$&$0.468$&$0.661$&$0.722$&$1.804$&$0.583$&$0.812$&$0.892$ \\ 
$6.300$&$1.436$&$0.510$&$0.727$&$0.799$&$1.937$&$0.637$&$0.889$&$0.984$ \\ 
$6.500$&$1.575$&$0.560$&$0.802$&$0.886$&$2.072$&$0.695$&$0.973$&$1.085$ \\ 
$6.700$&$1.725$&$0.615$&$0.884$&$0.982$&$2.202$&$0.755$&$1.063$&$1.192$ \\ 
$6.900$&$1.871$&$0.671$&$0.967$&$1.078$&$2.318$&$0.816$&$1.157$&$1.303$ \\ 
$7.100$&$1.993$&$0.721$&$1.040$&$1.165$&$2.415$&$0.875$&$1.251$&$1.414$ \\ 
$7.300$&$2.077$&$0.760$&$1.097$&$1.234$&$2.489$&$0.930$&$1.343$&$1.524$ \\ 
$7.500$&$2.120$&$0.788$&$1.138$&$1.282$&$2.540$&$0.979$&$1.432$&$1.630$ \\ 
$7.700$&$2.137$&$0.808$&$1.167$&$1.317$&$2.572$&$1.021$&$1.517$&$1.733$ \\ 
$7.900$&\nodata&\nodata&\nodata&\nodata&$2.593$&$1.056$&$1.598$&$1.833$ \\ 
$8.100$&\nodata&\nodata&\nodata&\nodata&$2.607$&$1.085$&$1.674$&$1.929$ \\ 
$8.300$&\nodata&\nodata&\nodata&\nodata&$2.620$&$1.107$&$1.744$&$2.021$ \\ 
$8.500$&\nodata&\nodata&\nodata&\nodata&$2.632$&$1.124$&$1.808$&$2.109$ \\ 
$8.700$&\nodata&\nodata&\nodata&\nodata&$2.643$&$1.135$&$1.866$&$2.194$ \\ 
$8.900$&\nodata&\nodata&\nodata&\nodata&$2.653$&$1.141$&$1.920$&$2.276$ \\ 
$9.100$&\nodata&\nodata&\nodata&\nodata&$2.659$&$1.141$&$1.976$&$2.358$ \\ 
$9.300$&\nodata&\nodata&\nodata&\nodata&$2.656$&$1.137$&$2.037$&$2.442$ \\ 
$9.500$&\nodata&\nodata&\nodata&\nodata&$2.642$&$1.128$&$2.108$&$2.530$ \\ 
$9.700$&\nodata&\nodata&\nodata&\nodata&$2.617$&$1.116$&$2.190$&$2.623$ \\ 
$9.900$&\nodata&\nodata&\nodata&\nodata&$2.585$&$1.102$&$2.278$&$2.718$ \\ 
\enddata
\end{deluxetable*}

%% file: tab3.tex
\begin{deluxetable*}{lccccc}
\tablecaption{Component Analysis at $4 < |Z| < 6$~kpc\label{tab:tab3}}
\tablehead{
   \colhead{} &
   \colhead{$\langle {\rm [Fe/H]} \rangle$} &
   \colhead{$\langle v_\phi \rangle$} &
   \colhead{$\sigma ({\rm [Fe/H]})$} &
   \colhead{$\sigma (v_\phi)$} &
   \colhead{Contribution} \\
   \colhead{Stellar Population} &
   \colhead{(dex)} &
   \colhead{(km~s$^{-1}$)} &
   \colhead{(dex)} &
   \colhead{(km~s$^{-1}$)} &
   \colhead{(\%)}
}
\startdata
Splash Disk (SD) &$ -0.57 \pm 0.03 $&$ 90.1 \pm 11.2 $&$ 0.31 \pm 0.01 $&$ 91.4 \pm 16.5 $&$ 24.8 \pm 4.9 $\\
Metal Weak Thick Disk (MWTD) &$ -0.90 \pm 0.17 $&$ 129.3 \pm 1.7 $&$ 0.40 \pm 0.02 $&$ 31.8 \pm 1.2 $&$ 5.7 \pm 1.1 $\\
{\it Gaia}-Sausage-Enceladus (GSE) &$ -1.16 \pm 0.04 $&$ -60.3 \pm 1.3 $&$ 0.20 \pm 0.04 $&$ 32.3 \pm 0.5 $&$ 0.5 \pm 0.1 $\\
Prograde Metal-rich Halo (MRH) &$ -1.58 \pm 0.05 $&$ 64.6 \pm 5.3 $&$ 0.46 \pm 0.08 $&$ 76.8 \pm 1.2 $&$ 30.3 \pm 11.8 $\\
Prograde Metal-poor Halo (MPH) &$ -2.50 \pm 0.00 $&$ 58.6 \pm 9.9 $&$ 0.54 \pm 0.05 $&$ 73.7 \pm 0.8 $&$ 35.2 \pm 5.7 $\\
Retrograde Halo Structure(s) (RHS) &$ -2.28 \pm 0.14 $&$ -142.0 \pm 11.4 $&$ 0.98 \pm 0.05 $&$ 54.7 \pm 1.5 $&$ 3.5 \pm 0.1 $\\
\enddata
\end{deluxetable*}